\pdfoutput=1
\PassOptionsToPackage{bookmarks=false,colorlinks=true,linkcolor=blue,urlcolor=cyan,filecolor=green,citecolor=magenta,pdfstartview=FitV,hyperfootnotes=false,pdftitle={Critical collapse remains discretely self-similar at large D},pdfauthor={Christian Ecker, Florian Ecker, Daniel Grumiller},pdfsubject={Critical collapse remains discretely self-similar at large D},pdfkeywords={discrete self-similarity, spacetime crystals, critical collapse, large D expansion, general relativity, cosmic censorship, exact solutions, naked singularities},bookmarksopen=true}{hyperref}
\documentclass[aps,prl,10pt,twocolumn,superscriptaddress,amsmath,amssymb]{revtex4-2}

\usepackage{graphicx,bm,braket,mathrsfs,orcidlink,pgfplots}
\usepackage[normalem]{ulem}
\usepackage{tikz}
\usepackage{booktabs}
\usetikzlibrary{decorations.pathmorphing,backgrounds,calc,shapes,arrows,shapes.misc,shadows}
\tikzset{cross/.style={cross out, draw=black, minimum size=2*(#1-\pgflinewidth), inner sep=0pt, outer sep=0pt}, cross/.default={1pt}}
\usepgfplotslibrary{fillbetween}
\pgfplotsset{compat=1.18}

\DeclareMathOperator{\extdm}{d}
\newcommand{\extd}{\extdm\!}
\newcommand{\dd}{\extd}

\begin{document}

%%%%%%%%%%%%%%%%%%%%%%%%%%%%%%%%
%%% TITLE, AUTHORS, ABSTRACT %%%
%%%%%%%%%%%%%%%%%%%%%%%%%%%%%%%%

\preprint{TUW-26-xx}

\def\mytitle{Critical collapse remains discretely self-similar at large $\boldsymbol{D}$}
\title{\mytitle}

\author{\orcidlink{0000-0002-8669-4300}Christian Ecker}
\email{ecker@itp.uni-frankfurt.de}
\affiliation{%
Institute for Theoretical Physics, Goethe University\\ 60438, Frankfurt am Main, Germany
}%

\author{\orcidlink{0000-0002-0449-0081}Florian Ecker}
\email{fecker@hep.itp.tuwien.ac.at}
\affiliation{%
Institute for Theoretical Physics, TU Wien\\
Wiedner Hauptstrasse 8–10/136, A-1040 Vienna, Austria
}\affiliation{%
Quantum Technology Laboratories GmbH\\ 
Clemens-Holzmeister-Strasse 6/6, A-1100 Vienna, Austria
}

\author{\orcidlink{0000-0001-7980-5394}Daniel Grumiller}
\email{grumil@hep.itp.tuwien.ac.at}
\affiliation{%
Institute for Theoretical Physics, TU Wien\\
Wiedner Hauptstrasse 8–10/136, A-1040 Vienna, Austria
}

\begin{abstract}
When matter collapses at the threshold of black hole formation spacetime typically has an emergent continuous or discrete symmetry, depending on the matter content. For the Einstein--Klein--Gordon model in any finite dimension $D>3$ the associated critical spacetime solutions exhibit discrete self-similarity, characterized by the echoing period $\Delta$. Combining analytical and numerical results with dimensions up to $D=200$ we predict $\Delta$ remains finite at infinite dimension $D$, contrary to a recent conjecture. We find the fit $\Delta\approx1.6439+10.37/D-\mathcal{O}(1/D^2)$ and provide an explicit large-$D$ model.
\end{abstract}
% character limit for abstract: 600; we have 582 characters including spaces

\maketitle

%%%%%%%%%%%%%%%%%%%%%
%%% BODY OF PAPER %%%
%%%%%%%%%%%%%%%%%%%%%

% word count limit: 3750

\section{Introduction}
\label{sec:1}

Critical phenomena analogous to second order phase transitions were discovered in gravitational collapse by Choptuik \cite{Choptuik:1993jv} who numerically evolved the Einstein--Klein--Gordon equations (we set $4\pi\,G=c=1$)
\begin{equation}
    R_{\mu\nu}=2\,(\partial_\mu\phi)(\partial_\nu\phi)\qquad\qquad \nabla^2\phi=0
    \label{eq:intro}
\end{equation}
assuming spherical symmetry. Smooth one-parameter families of initial data were considered, such that the evolution leads to black-hole formation for parameters $p>p_\ast$ and dispersion for $p<p_\ast$. At the threshold $p=p_\ast$, the evolution approaches a critical solution separating these two outcomes.

The critical solution has three remarkable properties: First, it is universal within a given model, i.e., its form is independent of the particular family of initial data. Second, it develops a pointlike naked singularity where the Ricci scalar diverges. Third, it has an emergent symmetry, namely discrete self-similarity (DSS), characterized by Choptuik's echoing period $\Delta$. See \cite{Gundlach:2007gc,Gundlach:2025yje} for reviews and \cite{Abrahams:1993wa,Evans:1994pj,Koike:1995jm} for pioneering developments. Critical gravitational collapse is relevant to numerical relativity \cite{Gundlach:2007gc,Gundlach:2025yje,Choptuik:1996yg,Bruegmann:2003aw,Bizon:2011gg,Bizon:2015pfa}, cosmic censorship \cite{Wald:1997wa,Lehner:2010pn}, quantum gravity \cite{Ayal:1997ab,Chiba:1997ex,Goswami:2005fu,Tomasevic:2025clf,Tomasevic:2025kqy}, and primordial black holes \cite{Niemeyer:1997mt,Vaskonen:2020lbd,DeLuca:2020agl}. 

Moreover, the critical solution governs the behavior of near-critical evolutions. In particular, the mass of black holes formed just above threshold obeys the scaling law $M\propto(p-p_\ast)^{(D-3)\gamma}$, with a universal exponent $\gamma$ \cite{Choptuik:1993jv} depending on the spacetime dimension $D$. This exponent can be obtained by linearizing around the critical solution and determining the growth rate of its single unstable mode \cite{Gundlach:1995kd}. Constructing the critical solution is therefore essential for predicting the behavior of near-critical gravitational collapse.

Until recently, critical DSS solutions were known only numerically. Recently, a large-$D$ expansion was used to construct analytic DSS solutions \cite{Ecker:2026akf}. Treating the spacetime dimension $D$ as a parameter and expanding the Einstein--Klein--Gordon equations \eqref{eq:intro} in $\epsilon=1/(D-1)$ exploits the simplifications that arise at large dimension. The inverse-dimension expansion has proved useful in numerous other gravitational contexts \cite{Emparan:2013moa,Emparan:2013xia,Emparan:2015gva,Emparan:2020inr}. However, the resulting DSS solutions have important limitations: the Ricci scalar and scalar field are not generically finite in the limit $\epsilon\to0$, and the associated two-mode toy model is consistent only for $D\geq52$, beyond the range currently explored numerically, $D<14$, \cite{Sorkin:2005vz,Bland:2005kk}. Furthermore, the conjecture $\lim_{\epsilon\to0}\Delta\to0$ \cite{Ecker:2026joz} may imply that the large-$D$ expansion requires a reorganization.

In this Letter, we address these issues by combining numerical and analytic methods. We construct critical DSS solutions up to $D=200$ and determine their echoing periods. We reorganize the large-$D$ expansion to ensure that the Ricci and matter scalar remain finite and nonzero in the infinite-dimensional limit. Our numerical results show that the echoing period approaches a finite limit, $\Delta\simeq1.6439$, ruling out the conjectured vanishing of $\Delta$. We also develop an improved large-$D$ model that captures the large-$D$ behavior of the numerical solutions.

\section{Einstein--Klein--Gordon equations}\label{sec:2}

The field equations \eqref{eq:intro} in $D$ spacetime dimensions and the associated DSS geometries are discussed in \cite{Ecker:2026joz}. To numerically implement them in that work, Mart\'in-Garc\'ia--Gundlach (GG) gauge was used \cite{Martin-Garcia:2003xgm}. Since the present solver relies on the computational infrastructure of Reiterer--Trubowitz (RT) we use their gauge \cite{Reiterer:2012hnr},
\begin{multline}
\dd s^2=e^{-2\zeta(t,\xi)}\Big(-(1-\xi ^2)\dd t^2 - \frac{2}{\mu}\xi\dd t\dd\xi +\frac{1}{\mu^2}\dd\xi^2\\
+\frac{\xi^2}{(\mu+Q(t,\xi)\xi^2)^2}\,\extd\Omega^2_{S^{D-2}}\Big)
\label{eq:RT1}
\end{multline}
where $\mu$ is the RT echoing parameter and $\extd\Omega^2_{S^{D-2}}$ is the metric of the round $(D-2)$-sphere. The smooth functions $\zeta$ and $Q$ depend arbitrarily on the radial coordinate $\xi$.

Under $t\to t+2\pi $ we fix $Q$ to be periodic with normalized period $2\pi $ while $\zeta $ is quasiperiodic, $\zeta(t+2\pi,\xi)=\zeta(t,\xi)+\Delta/2$ with $\Delta$ being Choptuik's echoing period. The scalar field is antiperiodic, $\phi(t+2\pi,\xi )=-\phi(t,\xi )$, and the loci $\xi=0$ and $\xi=1$ are, respectively, the center and the self-similar horizon (SSH). For fixing the coordinates completely one is still free to choose the origin of $t$ which we shall come back to in the numerical setup.

The geometric meaning of the RT echoing parameter $\mu$ is to rescale the null coordinates $u^\pm=-(1\pm\xi)\exp{(-{\mu}t)}$ between two adjacent fundamental domains, $u^\pm\to{u^\pm}\exp{(-2\pi\mu)}$. It is related to the Choptuik echoing period
\begin{equation}
    \Delta = \frac{4\pi\mu}{1-\langle\psi_{-1}^2\rangle} = 4\pi\mu - \int_0^{4\pi}\!\!\!\extd t\,\omega_3^-(t,0)
    \label{eq:angelinajolie}
\end{equation}
where $\langle\psi_{-1}^2\rangle$ is the period average of the null derivative of the scalar field along the SSH, evaluated at the SSH, see Eq.~(29) of \cite{Ecker:2026joz} for a definition in GG gauge; alternatively, in RT gauge we have $\psi_{-1}=\frac{4\pi}{\Delta\,\sqrt{D-2}}\,\partial_t\phi(t,1)$. 

Following RT we introduce variables
\begin{subequations}
\label{eq:omega_def}
\begin{align}
 \omega_1(t,\xi) &= \omega_2^+(t,\xi)+\omega_2^-(t,\xi) + 2\zeta(t,\xi)Q(t,\xi)\\
 \omega_2^\pm(t,\xi) &= \mathcal{D}^\pm\big(\zeta(t,\xi)+\ln(\mu+Q(t,\xi)\xi^2)\big)\mp\mu\\\omega_3^\pm(t,\xi) &= \mathcal{D}^\pm\zeta(t,\xi)\mp\mu\qquad\;\omega_4^\pm(t,\xi) = \mathcal{D}^\pm\phi(t,\xi) 
\end{align}
\end{subequations}
with null derivatives $\mathcal{D}^\pm=\pm\partial_t+\mu(1\pm\xi)\partial_\xi$. This definition implies $(\omega_1,\omega_2^\pm,\omega_3^\pm,\omega_4^\pm)\to(\omega_1,\omega_2^\pm,\omega_3^\pm,-\omega_4^\pm)$ under $t\to{t}+2\pi$. Together with the gauge choice \eqref{eq:RT1} the Einstein--Klein--Gordon equations \eqref{eq:intro} in arbitrary spacetime dimensions $D>3$ reduce to a set of equations $\Omega_i^\pm=0$ with $i=1,2,3,4,5$:
\begin{subequations}
\label{eq:eom}
\begin{align}
    & \Omega_1^\pm=\xi (\mathcal{D}^\pm \pm \mu)\,\omega_1+\mu(D-2)\omega_1+\frac{\xi}{4}\Big[
(D-3)\big(\omega_1^2\nonumber\\
&+(\omega_2^+)^2+(\omega_2^-)^2-2\omega_2^\mp(\omega_2^\pm+\omega_1)\big)-2(D-1)\omega_1\omega_2^\pm\nonumber\\
&+4\omega_3^\pm(\omega_1+\omega_2^\pm-\omega_2^\mp)-4(\omega_4^\pm)^2\Big]=0\\
&\Omega_2^\pm=\xi (\mathcal{D}^\pm \pm \mu)\,\omega_2^\mp + \mu \big[(D-3)\omega_1+\omega_2^++\omega_2^-\big]
\nonumber\\
&+\frac{\xi}{4}\Big[(D-3)\big(\omega_1^2+(\omega_2^+)^2+(\omega_2^-)^2-2\omega_1(\omega_2^++\omega_2^-)\big)\nonumber\\
&-2(D-1)\omega_2^+\omega_2^- \Big]=0\\
&\Omega_3^\pm=\xi (\mathcal{D}^\pm \pm \mu)\,\omega_3^\mp - \frac{\mu }{2}(D-3)(D-2)\omega_1\nonumber\\
&+ \frac{\xi}{4}\,(D-2)\Big[(D-3)\big((\omega_2^++\omega_2^-)\omega_1-\frac12\omega_1^2 \nonumber\\
&-\frac12(\omega_2^+-\omega_2^-)^2\big)-2\omega_4^+\omega_4^-\Big]=0\\
&\Omega_4^\pm=\xi (\mathcal{D}^\pm \pm \mu)\,\omega_4^\mp +\frac{\mu}{2}\,(D-2)\,\big(\omega_4^++\omega_4^-\big)\nonumber\\
&-\frac{\xi}{2}\,(D-2)\,\big(\omega_4^+\omega_2^-+\omega_4^-\omega_2^+\big)=0\\
&\Omega_5^\pm=\xi (\mathcal{D}^\pm\pm\mu)\,\omega_2^\pm+2\mu (\omega_2^\pm-\omega_3^\pm)+\xi\Big[2\omega_3^\pm\omega_2^\pm\nonumber\\
& -(\omega_2^\pm)^2-(\omega_4^\pm)^2\Big]=0 \label{eq:constraints}
\end{align}
\end{subequations} 
A key aspect of these equations is that they are only bilinear/quadratic in the variables $\omega_i$, which is exploited in the numerical implementation below. For $D=4$, Eqs.~\eqref{eq:eom} coincide precisely with Eqs.~(11) in \cite{Reiterer:2012hnr}, with $\Omega_5^\pm={\mathit\Omega}_{2\textrm{\#}}^\pm$. 

As explained by RT, we reformulate this system in terms of five independent equations on a parity extended two-dimensional domain that includes also negative radii, $\xi<0$. Introducing the parity operator $\mathcal{P}$ by the action $(t,\xi)\to(t,-\xi)$, we impose $\mathcal{P}\omega_1=-\omega_1$ and $\mathcal{P}\omega_i^\pm=-\omega_i^\mp$ for $i=2,3,4$. Having expressed all $\omega_i^+$ by $\omega_i^-$, it is sufficient to solve the $+$ versions of Eqs.~\eqref{eq:eom} for the variables $(\mu,\omega_1,\omega_2^-,\omega_3^-,\omega_4^-)$ on the extended domain $\xi\in[-1,1]$ to find a unique solution of the full Einstein--Klein--Gordon system for $\xi\geq 0$. Additionally, \eqref{eq:constraints} are constraints. The independent set used by our solver is, similarly to RT, given by $\big((1+\mathcal{P})\Omega_1,\,\Omega_2^+,\,\Omega_3^+,\,\Omega_4^+\big)$. Next, we summarize some details about the used algorithm~\footnote{%
We explain in the Supplemental Material (SM) the rationale behind our choices for the numerical setup. More details of our numerical results and the documented code will appear in a future publication.
}. 

\section{Numerical setup}\label{sec:3}

For solving the system above we use a spectral method on the domain $(t,\xi)\in[0,4\pi]\times[-1,1]$, expanding in a Fourier basis in $t$ and in a Chebyshev basis in $\xi$. The representation of a function $g(t,\xi)$ through its modes $g_{nm}$,
\begin{align}
    g(t,\xi)=\sum _{n,m\in \mathbb{Z}}g_{nm}\exp \big(\tfrac12\,in\,t+im\,\theta\big) \big\vert_{\theta=\arccos\xi} 
\end{align}
converts the field equations \eqref{eq:eom} into matrix equations. 

We solve for a finite set of these modes with $N_{\mathrm{Fourier}}=40$ and $N_{\mathrm{Chebyshev}}=80$ to approximate the critical solution, translating all non-linear terms in the equations to convolutions given by explicit bilinear operators in spectral space. This infrastructure is directly adopted from Sec.~3 in \cite{Reiterer:2012hnr}. The remaining translational freedom in $t$ is fixed by setting $\mathrm{Re}(\omega_4^-)_{1,0}=0$ and, starting from a sufficiently close initial guess, we use a Newton--Raphson method to converge on the critical solution. Taking the parity and periodicity conditions of the functions $(\omega_1,\omega_2^-,\omega_3^-,\omega_4^-)$ into account we obtain $40\cdot80\cdot2\cdot(\tfrac14+\tfrac12+\tfrac12+\tfrac12)=11200$ degrees of freedom where the first two factors are the mode numbers, the factor $2$ comes from real and imaginary parts, and the numbers in parentheses are symmetry factors of the functions $\omega_i$. The degree of freedom fixed by $t$-translations is balanced by the constant $\mu$. Typically, after four iterations the residual is brought down to $||\Omega^+_i||\sim10^{-25}$ where $||\cdot||$ is the weighted $\ell^1$-norm as defined by RT, $||g||=\sum_{n,m}\kappa_1^{|n|}\kappa_2^{|m|}(|\mathrm{Re}(g_{nm})|+|\mathrm{Im}(g_{nm})|)$ with $\kappa_1,\kappa_2>1$. Starting from $D=4$ we sequentially increase $D$ in steps of $1/2$, taking linear extrapolations of the two previous converged results as initial guesses. 

As discussed in Appendix B of \cite{Ecker:2026joz}, at successively larger values of $D$ a second (singular) solution branch approaches the critical solution. This renders the Jacobian close to singular and requires increased numerical precision to converge on the regular solution branch. In \cite{Ecker:2026joz}, this constrained the authors to $D\leq5.5$ with double precision arithmetic. The infrastructure  by RT allows surpassing this difficulty by using the \texttt{gmp} library~\cite{GMP} in C which allows storing dyadic numbers $x=a\cdot 2^b$ in terms of two arbitrary-precision integers $a$ and $b$. All internal arithmetic is performed with exact operations on these objects. Rounding only becomes necessary once the Jacobian is inverted at each Newton step, which we need to do with increasingly higher precision as $D$ grows. 

\section{Numerical results}\label{sec:4}

\begin{figure}[htb]
\centering
\includegraphics[width=\linewidth]{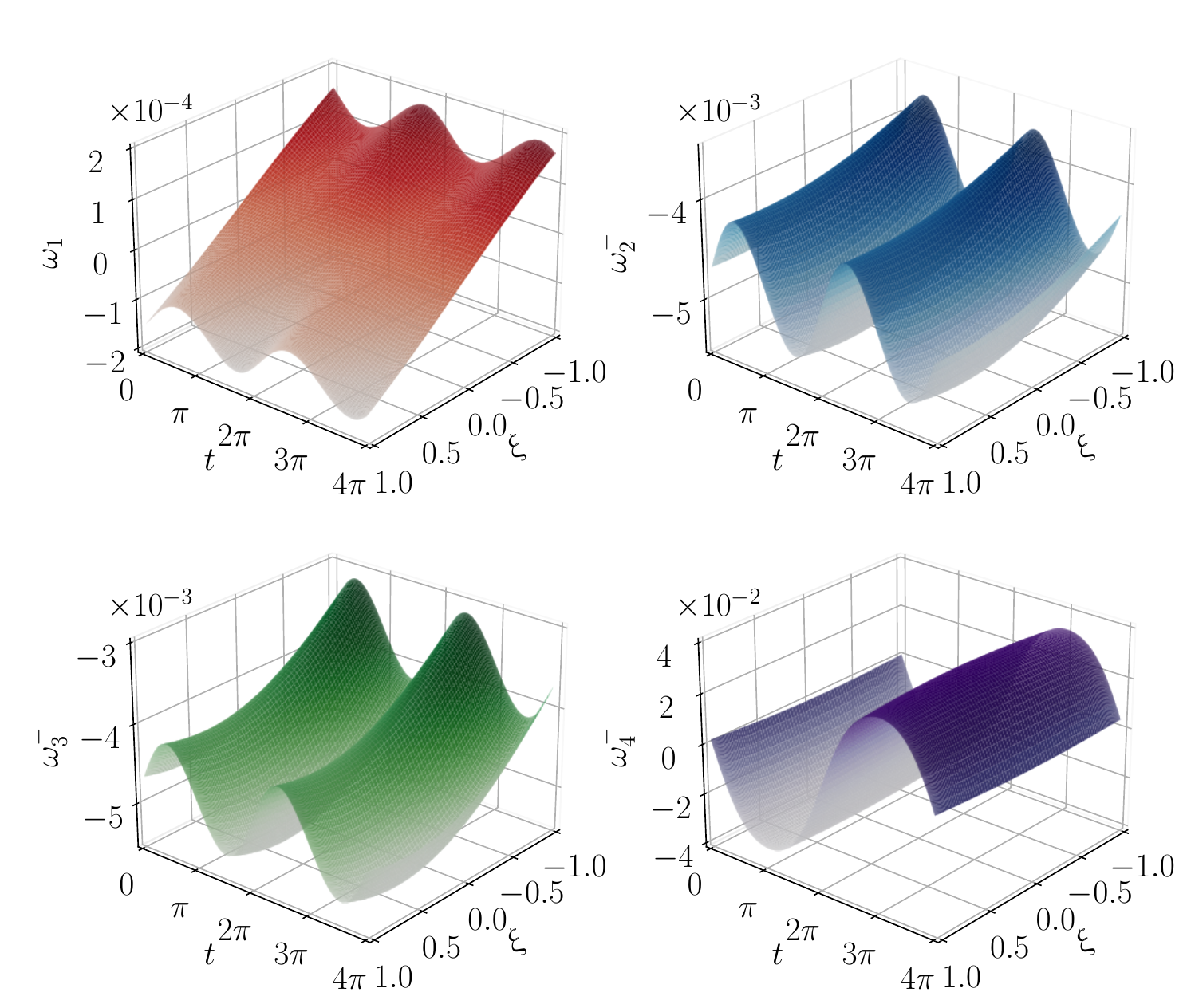}
\caption{Numerical solutions for the four fields $\omega_1$ (top left), $\omega_2^-$ (top right), $\omega_3^-$ (bottom left), and $\omega_4^-$ (bottom right) in $D=200$  dimensions as functions of $t$ and $\xi$.}
\label{fig:3}
\end{figure}

Using the numerical procedure discussed above, we construct critical DSS solutions up to $D=200$.
The numerical results for the physical fields $\omega_i$ are displayed in Fig.~\ref{fig:3}. 
The solution is visually dominated by the lowest-lying Fourier and Chebyshev modes, which matches with the converse observation \cite{Ecker:2026joz} that for very small dimensions $D\to3$ there is a lot of UV-drama, especially in the center. Evidently, this does not happen at large-$D$.

\begin{figure}[htb]
\centering
\includegraphics[width=\linewidth]{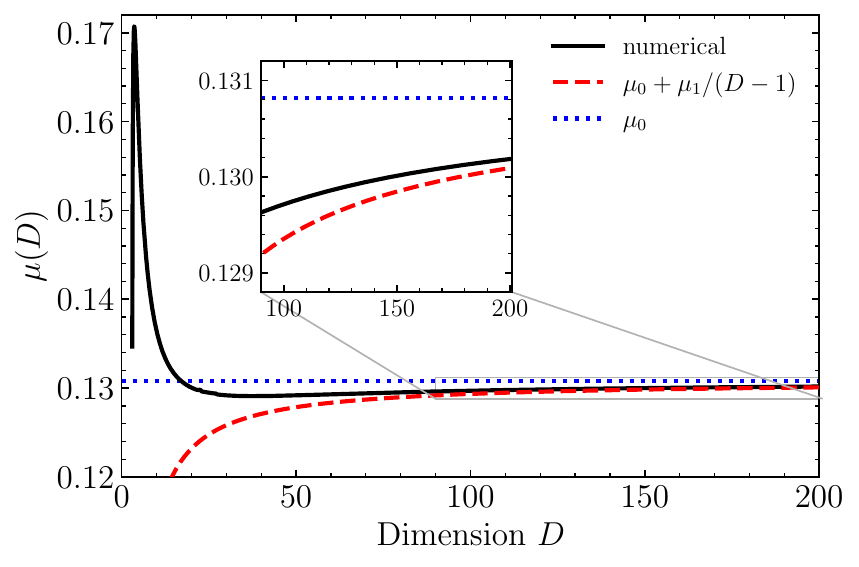}
\caption{Echoing parameter $\mu$. Black line: numerical results. Dashed/dotted lines: large-$D$ fits.}
\label{fig:1}
\end{figure}

Figure \ref{fig:1} displays the RT echoing parameter $\mu$, together with a large-$D$ fit $\mu=\mu_0+\mu_1\,\epsilon+\mathcal{O}(\epsilon^2)$ with $\epsilon=1/(D-1)$. Remarkably, the curve for $\mu$ flattens already around $D=20$ near its asymptotic value. 
We fit Taylor expansions $\mu=\sum_{n=0}^4\mu_n\epsilon^n$ to the numerical data over the range $160\leq D\leq 200$ from which we estimate 
\begin{equation}
\mu_0\approx 0.130818\qquad\qquad \mu_1\approx -0.1449\,.
\label{eq:mu}
\end{equation}

The data presented in Fig.~\ref{fig:1} are incompatible with the conjecture \cite{Ecker:2026joz} of vanishing $\Delta$ at large $D$ since $\mu$ quickly approaches a finite value. Combining the fit \eqref{eq:mu} with the relation \eqref{eq:angelinajolie} establishes the inequality
$\Delta\geq4\pi\mu=4\pi\mu_0+\mathcal{O}(\epsilon)\approx1.6439+\mathcal{O}(\epsilon)$. Thus, assuming we can trust the fit \eqref{eq:mu} we predict a lower bound for the Choptuik echoing period
\begin{equation}
\lim_{D\to\infty}\Delta \gtrsim 1.6439\,.
\label{eq:bound}
\end{equation}

\begin{figure}[htb]
\centering
\includegraphics[width=\linewidth]{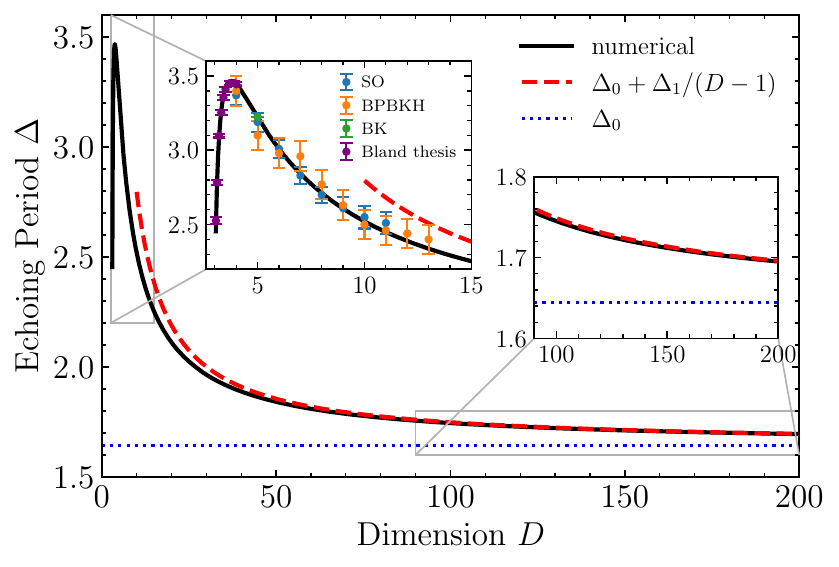}
\caption{Echoing period $\Delta$. Black line: numerical results. Dashed/dotted lines: large-$D$ fits. Other sources at low $D$: SO \cite{Sorkin:2005vz}, BPBKH \cite{Bland:2005kk}, BK \cite{Bland:2007sg}, and Bland's Ph.D.~thesis \cite{blandthesis}.}
\label{fig:2}
\end{figure}

Figure \ref{fig:2} displays our results for the Choptuik echoing period $\Delta$ up to $D=200$. The inset also displays earlier results from the literature for comparison. To see how close we are to the lower bound \eqref{eq:bound} we fit independently $\Delta=\sum_{n=0}^\infty\Delta_n\,\epsilon^n$ using our data, finding \footnote{%
A similar result was obtained independently by Nathan Moinvaziri, who used the AI Claude to determine the critical solution in dimensions up to $D=61$, see \href{https://github.com/nmoinvaz/choptuik-large-d}{\texttt{https://github.com/nmoinvaz/choptuik-large-d}}.
}
\begin{equation}
\Delta_0\approx1.6439\qquad\qquad\Delta_1\approx10.37\,.
\label{eq:delta}
\end{equation}

Comparing \eqref{eq:delta} with \eqref{eq:bound} reveals that within our accuracy we are saturating the bound.
This asks for an explanation, which we shall provide after reconsidering large-$D$ analytics.

\section{Refined large-$\boldsymbol{D}$ analytics}\label{sec:5}

Recently, we obtained analytic results at large $D$ \cite{Ecker:2026akf} and a perturbative $1/D$ expansion that allows to determine all observables order by order. Our solution depends on one free function $\beta(\tau)$, where $\tau$ is the GG time \footnote{%
See the SM for the coordinate transformation between RT and GG gauges.
}. In our previous work we were agnostic about the large $D$ scaling of $\beta$ (other than implicitly assuming it does not diverge too strongly with $D$). 

In this work, we propose the scaling
\begin{equation}
\beta(\tau) = \sqrt{\epsilon}\,\tilde\beta(\tau)\qquad\qquad \epsilon=\frac{1}{D-1}
\label{eq:beta}
\end{equation}
with an order-one function $\tilde\beta(\tau)$. The proposal \eqref{eq:beta} has a numerical and an analytical motivation.

The numerical motivation is that the scaling \eqref{eq:beta} explains why the bound \eqref{eq:bound} is saturated: For $\lim_{D\to\infty}\beta\to0$ the quantity $\psi_{-1}=\beta/\sqrt{1+\beta^2}+\mathcal{O}(\epsilon)$ vanishes and thus the exact relation \eqref{eq:angelinajolie} reduces to $\Delta=4\pi\mu+\mathcal{O}(\epsilon)$. Therefore, if there was an independent reason for \eqref{eq:beta} we would establish an explanation for the saturation of the bound \eqref{eq:bound} at infinite $D$.

The analytical motivation provides such a reason. As shown in \cite{Ecker:2026akf}, the Ricci scalar at large-$D$,
\begin{equation}
    R = -\frac{\beta^2\,e^{-2\tau}}{\epsilon\,\sqrt{1+\beta^2x^2}}\,\big(1+\mathcal{O}(\epsilon)\big) 
\label{eq:R}
\end{equation}
diverges in the limit $\epsilon\to0$. However, exactly for the scaling \eqref{eq:beta} we get a finite result for \eqref{eq:R}. Moreover, as shown in the SM to \cite{Ecker:2026akf} also the scalar field 
\begin{equation}
    \phi = \oint\extd\tau\,\frac{\beta}{\sqrt{\epsilon}\,\sqrt{1+\beta^2}}\,\big(1+\mathcal{O}(\epsilon)\big)
    \label{eq:phi}
\end{equation}
remains finite only for the scaling \eqref{eq:beta}. This provides sound geometric and physical reasons to insist on \eqref{eq:beta}.

The scaling \eqref{eq:beta} simplifies the analytics of \cite{Ecker:2026akf}. We provide here results for the first nontrivial terms of the metric functions $f,\Omega$ and the matter functions $\Pi,\Psi$ \footnote{%
The SM displays two additional subleading orders in $\epsilon$.
}:
\begin{align}
    f &= 1 - \frac{1-x^2}{2}\,\tilde\beta^2\,\epsilon
     & \Omega &= \tilde\beta^2\epsilon \label{eq:fpert} \\
    \Pi &= \sqrt{\epsilon}\,\tilde\beta & 
    \Psi &= \sqrt{\epsilon}\,\big(\tilde\beta+\tilde\beta'\big)\epsilon\label{eq:psipert}
\end{align}

Combining our analytical results with the large-$D$ fits \eqref{eq:mu}, \eqref{eq:delta} and the exact equality \eqref{eq:angelinajolie} establishes a relation between the L$^2$-norm of $\tilde\beta$ and the first few expansion coefficients,
\begin{equation}
\frac{\Delta_1-4\pi\,\mu_1}{4\pi\,\mu_0} = \frac{1}{2\pi}\,\int_0^{2\pi}\!\!\!\dd\theta\,\tilde\beta^2(\theta)\qquad\quad\theta=\frac{2\pi\tau}{\Delta_0}\,.
\end{equation}

The extrapolated large-$D$ scalar profile is well approximated by the following closed-form expression \footnote{%
The combination of analytics and numerics that leads to this approximation is discussed in the SM.
}, improving upon the two-mode model of Ref.~\cite{Ecker:2026akf}:
\begin{equation}
\label{eq:beta_largeD_fit}
\tilde\beta(\theta)=A_\ast\,\operatorname{Re}\bigg[\frac{e^{i\theta}}{1+q\,e^{i(2\theta+\delta)}}\bigg]
\end{equation}
The asymptotic expansions of $\mu$ and $\Delta$ independently fix the frequency, $\theta=\tau/(2\mu_0)$, and normalization, $A_\ast=[2(1-q^2)(\Delta_1-4\pi\mu_1)/(4\pi\mu_0)]^{1/2}\approx 3.8455$, leaving only two fitted parameters, $q\approx0.0551$ and $\delta\approx0.0985$. Equation \eqref{eq:beta_largeD_fit} is a Fourier expansion in odd harmonics, with successive mode amplitudes exponentially suppressed as $e^{-{\alpha}n}$, where $\alpha=-\ln{q}\approx2.90$.

\begin{figure}[htb]
\centering
\includegraphics[width=\linewidth]{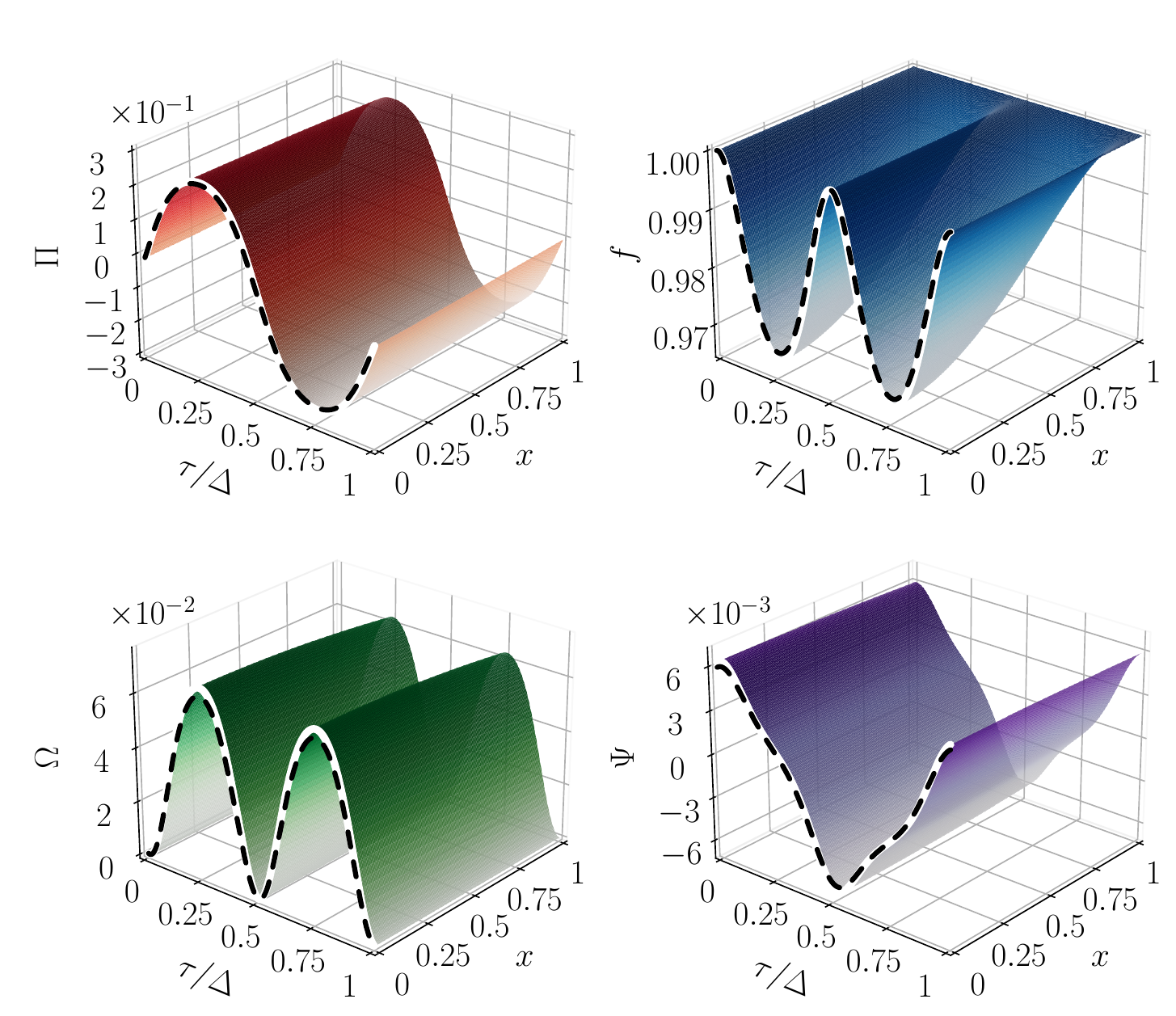}
\caption{Numerical results for $D=200$ from Fig.~\ref{fig:3} transformed to GG gauge and compared with large-$D$ approximation, to be contrasted with Fig.~2 of Ref.~\cite{Ecker:2026akf}. White solid lines: exact numerical profiles at $x=0$. Black dashed lines: corresponding large-$D$ predictions of Eq.~\eqref{eq:beta_largeD_fit}.}
\label{fig:4}
\end{figure}

Figure~\ref{fig:4} compares the numerical results for $D=200$, transformed from RT to GG gauge, with our perturbative predictions, demonstrating excellent agreement for $\Pi$, $f$, $\Omega$, and $\Psi$. The gauge transformation and perturbative calculations are detailed in the SM.

%%%%%%%%%%%%%%%
%%% SUMMARY %%%
%%%%%%%%%%%%%%%

\vspace*{-0.05truecm}

\section{Summary and outlook}\label{sec:8}

In this Letter, we numerically constructed DSS solutions of the Einstein--Klein--Gordon equations up to spacetime dimension $D=200$. Our results show that Choptuik's echoing period $\Delta$ approaches a finite, nonzero value in the infinite-dimensional limit \eqref{eq:delta}, contrary to the conjecture $\Delta\to0$ \cite{Ecker:2026joz}. This is surprising because it means that even at infinite dimension there is a pure number encoded in this theory, namely
\begin{equation}
    \Delta(D\to\infty) = \Delta_0 \approx 1.6439\,.
    \label{eq:conclusions}
\end{equation}
Thus, critical collapse remains DSS at large $D$.

We reorganized and simplified the perturbative large-$D$ expansion of \cite{Ecker:2026akf} such that Ricci and matter scalar remain finite at infinite $D$ by rescaling the free function $\beta$ arising in this construction, see \eqref{eq:beta}. Curiously, the leading order terms in the Ricci scalar \eqref{eq:R}, $R=-\tilde\beta^2\,e^{-2\tau}+\dots$, and the matter scalar \eqref{eq:phi}, $\phi=\oint\extd\tau\,\tilde\beta+\dots$, are homogeneous in space. This homogenuity does not correspond to an isometry of the metric to order $1/D$ and is broken by subleading corrections.

Moreover, we combined numerical and analytical results to provide an excellent large-$D$ approximation \eqref{eq:beta_largeD_fit} for the free function $\tilde\beta(2\pi\tau/\Delta_0)$ that depends on two free parameters that we fitted, a damping constant $q\approx0.0551$ and a phase $\delta\approx0.0985$.

Several questions naturally arise from these results. An outstanding analytical challenge is to derive large-$D$ numbers like \eqref{eq:conclusions} from first principles rather than extracting them from numerical fits. This will likely require to add instanton corrections of the form $e^{-1/\epsilon}$ to our power-law expansion in $\epsilon$. Additionally, it may be necessary to extend the solution beyond the SSH and to match with the Rozali--Way regime \cite{Rozali:2018yrv}. Perhaps some of the methods developed in the discussion of large-$D$ dynamics assuming continuous self-similarity \cite{Clark:2025tqi} can be applied to DSS as well, in particular, a separation into slow and fast dynamics at large $D$. It could also be rewarding to attempt a derivation of the large-$D$ approximation \eqref{eq:beta_largeD_fit} and the fitting parameters therein.

Given that we ruled out the conjecture $\Delta\to0$ at large $D$, it would be interesting to revisit the corresponding low-dimensional conjecture, namely $\Delta\to0$ as $D\to3^+$ \cite{Ecker:2026joz}. This will require substantially higher spectral resolution, in particular considerably more Fourier modes, since the solutions develop increasingly pronounced short-scale structure as the dimension decreases.

A further direction would be to extend the large-$D$ critical solution all the way to the Cauchy horizon and investigate its smoothness properties. In $D=4$, the corresponding analysis found the metric and scalar field to be $C^1$ but not $C^2$ \cite{Gundlach:1996eg,Martin-Garcia:2003xgm}. It would be interesting to determine whether this regularity persists in the large-$D$ limit, or whether the limiting solution exhibits different smoothness properties.

Another important observable is the Choptuik scaling exponent $\gamma$, which characterizes the black-hole mass near threshold, $M\propto(p-p_\ast)^{(D-3)\gamma}$. A recent determination using the AI system Claude~[36] suggests $\gamma\to1/2$ as $D\to\infty$. Although this result has not yet been verified by human analysis, the limiting value $\gamma=1/2$ is eminently reasonable \cite{Emparan:2020inr} and merits further investigation.

Finally, it would be worthwhile to explore how robust the large-$D$ analysis is under generalizations of the model. These include relaxing spherical symmetry, considering different matter fields and self-interactions, and studying other gravitational theories or matter couplings. Such extensions could establish which features of the large-$D$ critical dynamics are universal and which depend on the particular matter content and symmetries of the Einstein--Klein--Gordon system.

%%%%%%%%%%%%%%%%%%%%%%%
%%%                 %%%
%%% ACKNOWLEDGMENTS %%%
%%%                 %%%
%%%%%%%%%%%%%%%%%%%%%%%

\acknowledgments

\paragraph{Acknowledgments.} We thank Craig Clark for discussions and Nathan Moinvaziri for correspondence. CE acknowledges support by the DFG through the CRC-TR 211 ``Strong-interaction matter under extreme conditions'' -- project number 315477589 -- TRR 211. This work was supported by the Austrian Science Fund (FWF) [Grants DOI: \href{https://www.fwf.ac.at/en/research-radar/10.55776/P33789}{10.55776/ P33789}, \href{https://www.fwf.ac.at/en/research-radar/10.55776/P36619}{10.55776/P36619}, \href{https://www.fwf.ac.at/en/research-radar/10.55776/PAT1871425}{10.55776/PAT1871425}].  

%%%%%%%%%%%%%%%%%%
%%% REFERENCES %%%
%%%%%%%%%%%%%%%%%%

\bibliographystyle{fullsort}
\bibliography{review}

\providecommand{\href}[2]{#2}\begingroup\raggedright\begin{thebibliography}{10}

\bibitem{Choptuik:1993jv}
M.~W. Choptuik, ``Universality and scaling in gravitational collapse of a
  massless scalar field,'' {\em Phys. Rev. Lett.} {\bf 70} (1993)
9--12.
%%CITATION = PRLTA,70,9;%%.

\bibitem{Gundlach:2007gc}
C.~Gundlach and J.~M. Martin-Garcia, ``{Critical phenomena in gravitational
  collapse},'' {\em Living Rev. Rel.} {\bf 10} (2007) 5,
  \href{http://www.arXiv.org/abs/0711.4620}{{\tt 0711.4620}}.

\bibitem{Gundlach:2025yje}
C.~Gundlach, D.~Hilditch, and J.~M. Mart{\'\i}n-Garc{\'\i}a, ``{Critical
  Phenomena in Gravitational Collapse},''
  \href{http://www.arXiv.org/abs/2507.07636}{{\tt 2507.07636}}.

\bibitem{Abrahams:1993wa}
A.~M. Abrahams and C.~R. Evans, ``{Critical behavior and scaling in vacuum
  axisymmetric gravitational collapse},'' {\em Phys. Rev. Lett.} {\bf 70}
  (1993) 2980--2983.

\bibitem{Evans:1994pj}
C.~R. Evans and J.~S. Coleman, ``{Observation of critical phenomena and
  selfsimilarity in the gravitational collapse of radiation fluid},'' {\em
  Phys. Rev. Lett.} {\bf 72} (1994) 1782--1785,
  \href{http://www.arXiv.org/abs/gr-qc/9402041}{{\tt gr-qc/9402041}}.

\bibitem{Koike:1995jm}
T.~Koike, T.~Hara, and S.~Adachi, ``{Critical behavior in gravitational
  collapse of radiation fluid: A Renormalization group (linear perturbation)
  analysis},'' {\em Phys. Rev. Lett.} {\bf 74} (1995) 5170--5173,
  \href{http://www.arXiv.org/abs/gr-qc/9503007}{{\tt gr-qc/9503007}}.

\bibitem{Choptuik:1996yg}
M.~W. Choptuik, T.~Chmaj, and P.~Bizon, ``{Critical behavior in gravitational
  collapse of a Yang-Mills field},'' {\em Phys. Rev. Lett.} {\bf 77} (1996)
  424--427, \href{http://www.arXiv.org/abs/gr-qc/9603051}{{\tt gr-qc/9603051}}.

\bibitem{Bruegmann:2003aw}
B.~Bruegmann, W.~Tichy, and N.~Jansen, ``{Numerical simulation of orbiting
  black holes},'' {\em Phys. Rev. Lett.} {\bf 92} (2004) 211101,
  \href{http://www.arXiv.org/abs/gr-qc/0312112}{{\tt gr-qc/0312112}}.

\bibitem{Bizon:2011gg}
P.~Bizon and A.~Rostworowski, ``{On weakly turbulent instability of anti-de
  Sitter space},'' {\em Phys. Rev. Lett.} {\bf 107} (2011) 031102,
  \href{http://www.arXiv.org/abs/1104.3702}{{\tt 1104.3702}}.

\bibitem{Bizon:2015pfa}
P.~Bizo{\'n}, M.~Maliborski, and A.~Rostworowski, ``{Resonant Dynamics and the
  Instability of Anti{\textendash}de Sitter Spacetime},'' {\em Phys. Rev.
  Lett.} {\bf 115} (2015), no.~8, 081103,
  \href{http://www.arXiv.org/abs/1506.03519}{{\tt 1506.03519}}.

\bibitem{Wald:1997wa}
R.~M. Wald, ``{Gravitational collapse and cosmic censorship},'' pp.~69--85.
\newblock 10, 1997.
\newblock \href{http://www.arXiv.org/abs/gr-qc/9710068}{{\tt gr-qc/9710068}}.

\bibitem{Lehner:2010pn}
L.~Lehner and F.~Pretorius, ``{Black Strings, Low Viscosity Fluids, and
  Violation of Cosmic Censorship},'' {\em Phys. Rev. Lett.} {\bf 105} (2010)
  101102, \href{http://www.arXiv.org/abs/1006.5960}{{\tt 1006.5960}}.

\bibitem{Ayal:1997ab}
S.~Ayal and T.~Piran, ``{Spherical collapse of a massless scalar field with
  semiclassical corrections},'' {\em Phys. Rev. D} {\bf 56} (1997) 4768--4774,
  \href{http://www.arXiv.org/abs/gr-qc/9704027}{{\tt gr-qc/9704027}}.

\bibitem{Chiba:1997ex}
T.~Chiba and M.~Siino, ``Disappearance of black hole criticality in
  semiclassical general relativity,'' {\em Mod. Phys. Lett.} {\bf A12} (1997)
709--718.
%%CITATION = MPLAE,A12,709;%%.

\bibitem{Goswami:2005fu}
R.~Goswami, P.~S. Joshi, and P.~Singh, ``{Quantum evaporation of a naked
  singularity},'' {\em Phys. Rev. Lett.} {\bf 96} (2006) 031302,
  \href{http://www.arXiv.org/abs/gr-qc/0506129}{{\tt gr-qc/0506129}}.

\bibitem{Tomasevic:2025clf}
M.~Toma{\v{s}}evi{\'c} and C.-H. Wu, ``{Unveiling horizons in quantum critical
  collapse},'' {\em JHEP} {\bf 04} (2026) 151,
  \href{http://www.arXiv.org/abs/2509.03584}{{\tt 2509.03584}}.

\bibitem{Tomasevic:2025kqy}
M.~Toma{\v{s}}evi{\'c} and C.-H. Wu, ``{Quantum critical collapse abhors a
  naked singularity},'' {\em Phys. Rev. D} {\bf 113} (2026), no.~10, 104029,
  \href{http://www.arXiv.org/abs/2509.03587}{{\tt 2509.03587}}.

\bibitem{Niemeyer:1997mt}
J.~C. Niemeyer and K.~Jedamzik, ``{Near-critical gravitational collapse and the
  initial mass function of primordial black holes},'' {\em Phys. Rev. Lett.}
  {\bf 80} (1998) 5481--5484,
  \href{http://www.arXiv.org/abs/astro-ph/9709072}{{\tt astro-ph/9709072}}.

\bibitem{Vaskonen:2020lbd}
V.~Vaskonen and H.~Veerm{\"a}e, ``{Did NANOGrav see a signal from primordial
  black hole formation?},'' {\em Phys. Rev. Lett.} {\bf 126} (2021), no.~5,
  051303, \href{http://www.arXiv.org/abs/2009.07832}{{\tt 2009.07832}}.

\bibitem{DeLuca:2020agl}
V.~De~Luca, G.~Franciolini, and A.~Riotto, ``{NANOGrav Data Hints at Primordial
  Black Holes as Dark Matter},'' {\em Phys. Rev. Lett.} {\bf 126} (2021),
  no.~4, 041303, \href{http://www.arXiv.org/abs/2009.08268}{{\tt 2009.08268}}.

\bibitem{Gundlach:1995kd}
C.~Gundlach, ``{The Choptuik space-time as an eigenvalue problem},'' {\em Phys.
  Rev. Lett.} {\bf 75} (1995) 3214--3217,
  \href{http://www.arXiv.org/abs/gr-qc/9507054}{{\tt gr-qc/9507054}}.

\bibitem{Ecker:2026akf}
C.~Ecker, F.~Ecker, and D.~Grumiller, ``{Analytic Discrete Self-Similar
  Solutions of Einstein-Klein-Gordon at Large D},'' {\em Phys. Rev. Lett.} {\bf
  136} (2026), no.~19, 191401, \href{http://www.arXiv.org/abs/2601.14358}{{\tt
  2601.14358}}.

\bibitem{Emparan:2013moa}
R.~Emparan, R.~Suzuki, and K.~Tanabe, ``{The large D limit of General
  Relativity},'' {\em JHEP} {\bf 06} (2013) 009,
  \href{http://www.arXiv.org/abs/1302.6382}{{\tt 1302.6382}}.

\bibitem{Emparan:2013xia}
R.~Emparan, D.~Grumiller, and K.~Tanabe, ``{Large D gravity and low D
  strings},'' {\em Phys.Rev.Lett.} {\bf 110} (2013) 251102,
\href{http://www.arXiv.org/abs/1303.1995}{{\tt 1303.1995}}.
%%CITATION = ARXIV:1303.1995;%%.

\bibitem{Emparan:2015gva}
R.~Emparan, R.~Suzuki, and K.~Tanabe, ``{Evolution and End Point of the Black
  String Instability: Large D Solution},'' {\em Phys. Rev. Lett.} {\bf 115}
  (2015), no.~9, 091102, \href{http://www.arXiv.org/abs/1506.06772}{{\tt
  1506.06772}}.

\bibitem{Emparan:2020inr}
R.~Emparan and C.~P. Herzog, ``{Large D limit of Einstein's equations},'' {\em
  Rev. Mod. Phys.} {\bf 92} (2020), no.~4, 045005,
  \href{http://www.arXiv.org/abs/2003.11394}{{\tt 2003.11394}}.

\bibitem{Sorkin:2005vz}
E.~Sorkin and Y.~Oren, ``{On Choptuik's scaling in higher dimensions},'' {\em
  Phys. Rev.} {\bf D71} (2005) 124005,
\href{http://www.arXiv.org/abs/hep-th/0502034}{{\tt hep-th/0502034}}.
%%CITATION = HEP-TH 0502034;%%.

\bibitem{Bland:2005kk}
J.~Bland, B.~Preston, M.~Becker, G.~Kunstatter, and V.~Husain,
  ``{Dimension-dependence of the critical exponent in spherically symmetric
  gravitational collapse},'' {\em Class. Quant. Grav.} {\bf 22} (2005)
  5355--5364, \href{http://www.arXiv.org/abs/gr-qc/0507088}{{\tt
  gr-qc/0507088}}.

\bibitem{Ecker:2026joz}
C.~Ecker, F.~Ecker, D.~Grumiller, and T.~Jechtl, ``{Critical spacetime crystals
  in continuous dimensions},'' {\em SciPost Phys.} {\bf 20} (2026) 176,
  \href{http://www.arXiv.org/abs/2602.10185}{{\tt 2602.10185}}.

\bibitem{Martin-Garcia:2003xgm}
J.~M. Martin-Garcia and C.~Gundlach, ``{Global structure of Choptuik's critical
  solution in scalar field collapse},'' {\em Phys. Rev. D} {\bf 68} (2003)
  024011, \href{http://www.arXiv.org/abs/gr-qc/0304070}{{\tt gr-qc/0304070}}.

\bibitem{Reiterer:2012hnr}
M.~Reiterer and E.~Trubowitz, ``{Choptuik's critical spacetime exists},'' {\em
  Commun. Math. Phys.} {\bf 368} (2019), no.~1, 143--186,
  \href{http://www.arXiv.org/abs/1203.3766}{{\tt 1203.3766}}.

\bibitem{Note1}
We explain in the Supplemental Material (SM) the rationale behind our choices
  for the numerical setup. More details of our numerical results and the
  documented code will appear in a future publication.

\bibitem{GMP}
T.~Granlund and {the GMP development team}, {\em {GNU MP}: The {GNU} Multiple
  Precision Arithmetic Library}, 6.3.0~ed., 2023.

\bibitem{Bland:2007sg}
J.~Bland and G.~Kunstatter, ``{The 5-D Choptuik critical exponent and
  holography},'' {\em Phys. Rev. D} {\bf 75} (2007) 101501,
  \href{http://www.arXiv.org/abs/hep-th/0702226}{{\tt hep-th/0702226}}.

\bibitem{blandthesis}
J.~B. Bland, {\em Dimension Dependence of the Critical Phenomena in
  Gravitational Collapse of Massless Scalar Field}.
\newblock PhD thesis, University of Manitoba, 2006.

\bibitem{Note2}
A similar result was obtained independently by Nathan Moinvaziri, who used the
  AI Claude to determine the critical solution in dimensions up to $D=61$, see
  \protect \href {https://github.com/nmoinvaz/choptuik-large-d}{\protect
  \texttt {https://github.com/nmoinvaz/choptuik-large-d}}.

\bibitem{Note3}
See the SM for the coordinate transformation between RT and GG gauges.

\bibitem{Note4}
The SM displays two additional subleading orders in $\epsilon $.

\bibitem{Note5}
The combination of analytics and numerics that leads to this approximation is
  discussed in the SM.

\bibitem{Rozali:2018yrv}
M.~Rozali and B.~Way, ``{Gravitating scalar stars in the large D limit},'' {\em
  JHEP} {\bf 11} (2018) 106, \href{http://www.arXiv.org/abs/1807.10283}{{\tt
  1807.10283}}.

\bibitem{Clark:2025tqi}
C.~R. Clark and G.~L. Pimentel, ``{Black hole critical collapse in infinite
  dimensions: continuous self-similar solutions},'' {\em JHEP} {\bf 11} (2025)
  017, \href{http://www.arXiv.org/abs/2504.10590}{{\tt 2504.10590}}.

\bibitem{Gundlach:1996eg}
C.~Gundlach, ``{Understanding critical collapse of a scalar field},'' {\em
  Phys. Rev. D} {\bf 55} (1997) 695--713,
  \href{http://www.arXiv.org/abs/gr-qc/9604019}{{\tt gr-qc/9604019}}.

\end{thebibliography}\endgroup

%%%%%%%%%%%%%%%%%%%%%%%%%%%%%
%%%                       %%%
%%% SUPPLEMENTAL MATERIAL %%%
%%%                       %%%
%%%%%%%%%%%%%%%%%%%%%%%%%%%%%

\clearpage
\onecolumngrid
\setcounter{equation}{0}
\renewcommand{\theequation}{S\arabic{equation}}

\section{Supplemental Material}

In this Supplemental Material (SM), we elaborate on some technical aspects pertaining to the main text. While none of the material in this SM is crucial, it can be helpful for explicit checks, for pushing the large-$D$ perturbation theory to higher orders, or for refinements of our explicit large-$D$ model presented in the main text.

This SM is organized as follows. In the first section, we explain our choice of the numerical scheme that was used to produce the numerical results discussed in the main text. In the second section, we transform from Reiterer--Trubowitz to Mart\'in-Garc\'ia--Gundlach gauge. In the third section, we list the first three non-trivial orders in the large-$D$ perturbation theory. In the fourth section, we construct an explicit large-$D$ model that is exploited in the main text. On the final page, we display in Table \ref{tab:1} our numerical results of the echoing period in dimensions between $3.046$ and $200$; for comparison we added as a final entry our best-fit result for $D\to\infty$. The results for $D>28$ have sufficient accuracy so that we trust all displayed digits, and we checked that varying the number of Fourier or Chebyshev modes does not alter these digits. The results for $D\leq5.5$ contain also data from our previous work \cite{Ecker:2026joz}.

\subsection{Choice of numerical scheme}

Choptuik's original code evolved one-parameter families of initial data in time and checked whether or not there is an apparent horizon to decide if a black hole has formed. Using this code, he discovered the echoing period in $D=4$ is approximately given by
\begin{equation}
    \Delta_{\textrm{\tiny Ch}} = 3.44\pm 10^{-2}\,.
\end{equation}
Variants of this approach were subsequently used by Sorkin and Oren \cite{Sorkin:2005vz} and Bland et al.~\cite{Bland:2005kk,blandthesis} to study critical collapse in higher dimensions, obtaining echoing periods up to $D=13$. One limiting factor of these codes is that gradients become very steep in higher dimensions \cite{Emparan:2013moa,Emparan:2020inr}. Another inefficient aspect is that one needs to construct many spacetimes and only asymptotically approaches the critical DSS solution.

Gundlach circumvented this inefficiency by constructing the critical DSS solution directly, thereby reducing the problem to a nonlinear eigenvalue problem in which the echoing period $\Delta$ is an unknown to be determined \cite{Gundlach:1995kd}. As explained in great detail in \cite{Martin-Garcia:2003xgm}, it is sufficient to construct numerically a solution for a fundamental domain in the past patch, i.e., a trapezoid region bounded in space by the center and the SSH, and bounded in time by one period $\Delta$. They used spectral methods in time, i.e., a Fourier decomposition, and shooting methods in space. This code allowed a more precise determination of the echoing period,
\begin{equation}
    \Delta_{\textrm{\tiny GG}} = 3.445452402\pm10^{-9}\,.
\end{equation}
Generalizing their code to arbitrary dimensions $D$, Ecker et al.~obtained results for echoing periods both in lower and higher dimensions \cite{Ecker:2026joz}. However, a crucial limiting factor was identified in their Appendix: besides the regular DSS solution, the equations also have a solution that becomes singular at the SSH. At large $D$, the singular branch becomes increasingly difficult to distinguish numerically from the regular one. Near the SSH at $x=1$, its leading singular contribution scales as $(1-x)^{(D-4+f(D))/(2-f(D))}$, where $f(D)$ is slowly varying and tends to zero at large $D$. Since the exponent grows approximately as $D/2$, distinguishing the two branches by a Taylor expansion requires a number of terms that grows with $D/2$, making the calculation increasingly expensive. Mainly for this reason, \cite{Ecker:2026joz} did not progress beyond $D=5.5$.

Reiterer and Trubowitz used the same idea of Gundlach to solve a nonlinear eigenvalue problem. A key difference is their use of spectral methods in space in addition to a Fourier decomposition in time. Expanding all functions in Fourier and Chebyshev modes (as stated in the main text) converts the Einstein--Klein--Gordon equations into a set of nonlinear matrix equations. With their choice of variables, the equations are at most quadratic in the unknowns, so the linearized operator depends linearly on the background solution. This feature adds to the numerical stability of their scheme. Additional advantages, such as arbitrary precision integer arithmetics and a judiciously chosen split of the spectrum into UV and IR modes that are treated differently, are explained in detail in their original paper \cite{Reiterer:2012hnr}. While the main task of their work was to prove that Choptuik's critical DSS spacetime exists, a byproduct of their analysis is unprecedented precision in the determination of Choptuik's echoing period,
\begin{equation}    
\Delta_{\textrm{\tiny RT}}=3.44545240222782134997151194557638372444206917611562177269576807141080542911297710\pm 10^{-80}\,.
\end{equation}

The precision achieved by RT in $D=4$ motivated us to adopt part of their infrastructure in our solver to reach our goal of finding critical DSS solutions and the associated echoing periods in large $D$.

\subsection{Transformation from RT to GG gauge}
\label{app:RTtoGG}

To compare our numerical results obtained in RT gauge with the analytical large-$D$ expansion derived in GG gauge~\cite{Ecker:2026akf}, we construct the coordinate transformation between the two gauges and derive the corresponding relations between their field variables.
The GG metric is given by
\begin{equation}
\label{eq:GGmetric}
\dd s^2=e^{-2\tau}\left[a^2\left((x^2-f^2)\dd\tau^2-2x\,\dd\tau\dd x+\dd x^2\right)+x^2\extd\Omega^2_{S^{D-2}}\right]
\end{equation}
where $a=a(\tau,x)$ and $f=f(\tau,x)$.
It is related to the diagonal metric
\begin{equation}
\label{eq:polarareal}
\dd s^2=-\alpha^2(T,r)\dd T^2
+a^2(T,r)\dd r^2
+r^2\extd\Omega^2_{S^{D-2}}
\end{equation}
through the coordinate and field redefinitions
\begin{equation}
\label{eq:GGdefinitions}
T=-e^{-\tau}\qquad\qquad
r=e^{-\tau}x\qquad\qquad
f=\frac{\alpha}{a}\qquad\qquad\Omega=\frac{D-3}{x^2}(a^2-1)\,.
\end{equation}

In RT gauge, the line element reads
\begin{equation}
\label{eq:RTmetric}
\dd s^2=e^{-2\zeta(t,\xi)}
\left[
-(1-\xi^2)\dd t^2
-\frac{2\xi}{\mu}\dd t\dd\xi
+\frac{\dd\xi^2}{\mu^2}
+\frac{\xi^2}{(\mu+Q(t,\xi)\xi^2)^2}
\extd\Omega^2_{S^{D-2}}
\right]\,.
\end{equation}
Equating the angular components of \eqref{eq:GGmetric} and \eqref{eq:RTmetric}
identifies the areal radius,
\begin{equation}
\label{eq:arealradius}
r(t,\xi)=e^{-\tau(t,\xi)}x(t,\xi)
=\xi e^{-W(t,\xi)}\qquad\qquad W(t,\xi)=\zeta(t,\xi)+\ln H(t,\xi)\qquad\qquad H(t,\xi)=\mu+Q(t,\xi)\xi^2\,.
\end{equation}

Rather than determining the two coordinate functions
$\tau(t,\xi)$ and $x(t,\xi)$ independently, we introduce
a single function $B(t,\xi)$ through
%\begin{equation}
%\label{eq:Bdefinition}
$x(t,\xi)=\xi B(t,\xi)$.
%\end{equation}
Equation~\eqref{eq:arealradius} then immediately yields
\begin{equation}
\label{eq:tautransformation}
\tau(t,\xi)=W(t,\xi)+\ln B(t,\xi)\,.
\end{equation}
Since the SSH is located at $\xi=1$ in RT gauge and $x=1$ in GG gauge, the corresponding boundary conditions are
\begin{equation}
\label{eq:Bboundary}
B(t,1)=1\qquad\qquad \tau(t,1)=W(t,1)\,.
\end{equation}

The remaining function $B$ is determined by imposing the defining property of polar-areal coordinates, namely the absence of a mixed $rT$ metric component $g^{\mu\nu}\partial_\mu{T}\partial_\nu{r}=0$.
Using the RT null derivatives $\mathcal{D}^\pm=\pm\partial_t+\mu(1\pm\xi)\partial_\xi$, we can express this condition as
\begin{equation}
\label{eq:orthogonalityRT}
(\mathcal{D}^+T)(\mathcal{D}^-r)+(\mathcal{D}^-T)(\mathcal{D}^+r)=0\,.
\end{equation}

To evaluate these derivatives, we use the RT field definitions $\omega_2^\pm=\mathcal{D}^\pm W\mp\mu$\,.
Together with \eqref{eq:arealradius} and \eqref{eq:tautransformation}, this gives
\begin{equation}
\label{eq:RTderivatives}
\mathcal{D}^\pm r=e^{-W}A^\pm\qquad\qquad \mathcal{D}^\pm T=\frac{e^{-W}}{B}\left(\omega_2^\pm\pm\mu+\mathcal{D}^\pm\ln B \right)
\end{equation}
where $A^\pm(t,\xi)=\mu-\xi\omega_2^\pm(t,\xi)$.
Substituting into \eqref{eq:orthogonalityRT} and collecting derivatives of $\ln B$, we obtain
\begin{equation}
\label{eq:RTtoGG_PDE}
J\,\partial_t\ln B+K\,\partial_\xi\ln B=S
\end{equation}
with
\begin{equation}
\label{eq:JKS}
J=\xi(\omega_2^+-\omega_2^-)\qquad 
K=2\mu^2-\mu\xi \left[(1-\xi)\omega_2^++(1+\xi)\omega_2^-\right]\qquad
S=2\xi\omega_2^+\omega_2^--\mu\left[(1+\xi)\omega_2^++(1-\xi)\omega_2^-\right]\,.
\end{equation}
This first-order linear PDE for $\ln B$ completely determines the coordinate transformation and depends only on the RT echoing parameter $\mu$ and the numerical fields $\omega_2^\pm$.

We solve Eq.~\eqref{eq:RTtoGG_PDE} using the method of characteristics.
Parameterizing the characteristic curves by $t=t(s)$ and $\xi=s$, the total derivative along
a characteristic is
\begin{equation}
\frac{\dd\ln B}{\dd s}=\partial_t\ln B\frac{\dd t}{\dd s}+\partial_\xi\ln B\,.
\end{equation}
Choosing $\dd t/\dd s=J/K$, Eq.~\eqref{eq:RTtoGG_PDE} reduces to
\begin{equation}
\label{eq:characteristics}
\frac{\dd t}{\dd s}=\frac{J}{K}\qquad\qquad \frac{\dd\ln B}{\dd s}=\frac{S}{K}\,.
\end{equation}
The GG time coordinate follows from Eq.~\eqref{eq:tautransformation},
\begin{equation}
\label{eq:tauODE}
\frac{\dd\tau}{\dd s}=\partial_t W\frac{J}{K}+\partial_\xi W+\frac{S}{K}
\end{equation}
where the derivatives of $W$ can be expressed directly in terms of the numerical RT fields,
\begin{align}
\label{eq:Wderivatives}
\partial_t W=\mu+\frac{(1-\xi)\omega_2^+-(1+\xi)\omega_2^-}{2}\qquad\qquad \partial_\xi W=\frac{\omega_2^++\omega_2^-}{2\mu}\,.
\end{align}
We thus obtain a closed system of three ODEs,
\begin{equation}\label{eq:RTtoGG_ODEsystem}
\frac{\dd t}{\dd s}=\frac{J}{K}\qquad\qquad
\frac{\dd\ln B}{\dd s}=\frac{S}{K}\qquad\qquad
\frac{\dd\tau}{\dd s}=\partial_t W\frac{J}{K}+\partial_\xi W+\frac{S}{K}\,.
\end{equation}

For each initial RT time $t_{\textrm{\tiny SSH}}$, we integrate
this system inward from the SSH at $s=1$ to the
regular center at $s=0$, subject to the boundary conditions
\begin{equation}
\label{eq:characteristicBC}
t(1)=t_{\textrm{\tiny SSH}}\qquad\qquad
\ln B(1)=0\qquad\qquad
\tau(1)=\tau_{\textrm{\tiny SSH}}(t_{\textrm{\tiny SSH}})\,.
\end{equation}
The initial value of the GG time coordinate follows
from Eq.~\eqref{eq:Bboundary}. In particular,
Eq.~\eqref{eq:Wderivatives} implies
\begin{equation}
\label{eq:tauSSH}
\frac{\dd\tau_{\textrm{\tiny SSH}}}{\dd t}
=\mu-\omega_2^-(t,1),
\end{equation}
which can be integrated to give
\begin{equation}
\label{eq:tauSSHintegral}
\tau_{\textrm{\tiny SSH}}(t)=
\mu\,t-\int_0^t\omega_2^-(t',1)\dd t'
\end{equation}
where we fixed the remaining time-translation freedom
by choosing $\tau_{\textrm{\tiny SSH}}(0)=0$.

The quasi-periodicity of the coordinate transformation,
$\tau(t+2\pi,\xi)=\tau(t,\xi)+\Delta/2$,
provides an independent expression for the echoing period,
\begin{equation}
\label{eq:DeltaSSH}
\Delta=4\pi\mu-
\int_0^{4\pi}\omega_2^-(t,1)\dd t\,.
\end{equation}
This agrees with the corresponding relation obtained
from the RT fields at the center,
\begin{equation}
\Delta=4\pi\mu-
\int_0^{4\pi}\omega_3^-(t,0)\dd t
\end{equation}
and provides a useful consistency check.
The identity between the two integrals is a consequence of periodicity of $Q(t+2\pi,\xi)=Q(t,\xi)$, quasi-periodicity of $\zeta(t+2\pi,\xi)=\zeta(t,\xi)+\frac{\Delta}{2}$ and regularity of $\partial_\xi\zeta(t,0)=0$ at the center.
At the SSH $\omega_2^-(t,1)=\mu-\partial_t\left[\zeta(t,1)+\ln(\mu+Q(t,1))\right]$, where at the center we have $\omega_3^-(t,0)=\mu-\partial_t\zeta(t,0)$.
The difference between these expressions becomes a total derivative
%\begin{equation}
$\omega_2^-(t,1)-\omega_3^-(t,0)=\partial_t F(t)$ with $F(t)=\zeta(t,1)-\zeta(t,0)+\ln(\mu+Q(t,1))$.
%\end{equation}
Since $F(t+2\pi)=F(t)$, the corresponding integral vanishes
%\begin{equation}
$\int_0^{4\pi}\left[\omega_2^-(t,1)-\omega_3^-(t,0)\right]\dd t=-\int_0^{4\pi}\partial_t F(t)=-\left[F(4\pi)-F(0)\right]=0$,
%\end{equation}
which establishes the identity $\int_0^{4\pi}\omega_2^-(t,1)\dd t=\int_0^{4\pi}\omega_3^-(t,0)\dd t$.

In practice, we reconstruct the RT fields from their
Fourier--Chebyshev coefficients and evaluate them directly
at arbitrary points along the characteristics.
The plus fields are obtained from the parity relations
%\begin{equation}
$\omega_i^+(t,\xi)=-\omega_i^-(t,-\xi)$, $i=2,3,4$.
%\end{equation}
We then integrate Eq.~\eqref{eq:RTtoGG_ODEsystem}
using an adaptive ODE solver for a set of initial
times covering one complete RT period.
The resulting coordinate transformation
$(t,\xi)\mapsto(\tau,x)$ is used to reconstruct the
GG fields, which are subsequently interpolated onto
a rectangular grid in $(\tau,x)$.

Having determined the coordinate transformation, we reconstruct the four GG fields $\Pi$, $\Psi$, $\Omega$, and $f$ from the RT solution.
The GG matter fields are defined by
\begin{equation}
\label{eq:GGmatterdefinitions}
\Pi=\frac{1}{\sqrt{D-2}\,f}(\partial_\tau+x\partial_x)\psi\qquad\qquad \Psi=\frac{1}{\sqrt{D-2}\,x}\partial_x\psi
\end{equation}
where $\psi$ denotes the scalar field in the GG normalization. The RT matter variables are $\omega_4^\pm=D_\pm\varphi_{\rm num}$, where $\varphi_{\rm num}$ denotes the scalar variable represented by the numerical spectral coefficients.
Introducing the normalization factor
%\begin{equation}
$\mathcal C=
\frac{1}{\sqrt{D-2}}
\frac{\dd\psi}{\dd\varphi_{\rm num}}$,
%\end{equation}
the transformation follows from the identities
%\begin{equation}
$\partial_\tau+x\partial_x
=e^{-\tau}\partial_T$ and
$\partial_x=e^{-\tau}\partial_r$.
%\end{equation}
Combining these with the RT null derivatives gives
\begin{equation}
\Pi(\tau,x)=\frac{\mathcal C}{2B}\left(\frac{\omega_4^+}{A_+}-\frac{\omega_4^-}{A_-}\right)\qquad\qquad
\Psi(\tau,x)=\frac{\mathcal C}{2\xi B^2}
\left(\frac{\omega_4^+}{A_+}+\frac{\omega_4^-}{A_-}\right)\,.
\label{eq:PsiRTtoGG}
\end{equation}
For the numerical spectral coefficients employed here,
the normalization is $\mathcal C=1$.
Thus, both GG matter fields are obtained directly from
the numerical RT variables without explicitly
reconstructing the scalar field.

The transformation of the metric function $\Omega$ follows from the norm of the areal-radius gradient,
%\begin{equation}
$a^{-2}=g^{\mu\nu}\partial_\mu r\partial_\nu r=\frac{A_+A_-}{H^2}$.
%\end{equation}
Using the definition of $\Omega$ in Eq.~\eqref{eq:GGdefinitions}, we obtain
\begin{equation}
\Omega(\tau,x)=\frac{D-3}{\xi^2B^2}\bigg(\frac{H^2}{A_+A_-}-1\bigg)\,.
\label{eq:OmegaRTtoGG}
\end{equation}
The remaining function $H$ can be reconstructed from
the RT variables using
%\begin{equation}
$\partial_\xi\ln H=\frac{\omega_2^++\omega_2^- - \omega_3^+-\omega_3^-}{2\mu}$
%\end{equation}
together with $H(t,0)=\mu$. Consequently,
\begin{equation}
\label{eq:Hreconstruction}
H(t,\xi)=\mu\exp\left[
\int_0^\xi
\frac{\omega_2^++\omega_2^-
-\omega_3^+-\omega_3^-}{2\mu}
\,\dd s
\right]\,.
\end{equation}
Finally, the GG metric function $f$ follows from the exact radial field equation $\partial_x\ln f=x\,\Omega$. Imposing the SSH boundary condition $f(\tau,1)=1$ gives
\begin{equation}
f(\tau,x)=\exp\left[-\int_x^1s\,\Omega(\tau,s)\,\dd s\right]\,.
\label{eq:fRTtoGG}
\end{equation}
The integration is performed at fixed GG time $\tau$,
after interpolating $\Omega$ onto the rectangular
GG grid.

The expressions for $\Psi$ and $\Omega$ contain apparent singularities at $\xi=0$, which are removed by evaluating their regular center limits. In particular, one finds
\begin{align}
\label{eq:GGcenterlimits}
\Pi(\tau(t,0),0) &=-\frac{\mathcal C\,\omega_4^-(t,0)}
{\mu B(t,0)} \\
\Psi(\tau(t,0),0) &=\frac{\mathcal C}{B(t,0)^2}
\left[\frac{\partial_\xi\omega_4^-(t,0)}{\mu}
+\frac{\omega_4^-(t,0)\omega_2^-(t,0)}{\mu^2}
\right] \\
\Omega(\tau(t,0),0) &=\frac{D-3}{B(t,0)^2}\left[\frac{3\partial_\xi\omega_2^-(t,0)-\partial_\xi\omega_3^-(t,0)}{\mu}+\frac{\omega_2^-(t,0)^2}{\mu^2}\right]\,.
\end{align}
An independent check of the transformation is provided by the exact GG center constraint
\begin{equation}
\label{eq:GGcenterconstraint}
\Omega(\tau,0)=\frac{D-3}{D-1}\,\Pi(\tau,0)^2\,.
\end{equation}
Together with the independent determinations of $\Delta$ in Eq.~\eqref{eq:DeltaSSH}, this provides a test of the coordinate transformation and the normalization of the matter fields.

At large $D$, the reconstruction of $\Omega$ in Eq.~\eqref{eq:OmegaRTtoGG} involves the subtraction of nearly equal quantities. 
To reduce numerical cancellation, we evaluate this expression using
%\begin{equation}
$\Omega=\frac{D-3}{x^2}\left(e^\delta-1\right)$ and $\delta=\ln\frac{H^2}{A_+A_-}$
%\end{equation}
with a numerically stable evaluation of $e^\delta-1$.

The coordinate and field transformations presented above are exact at finite $D$. The large-$D$ expansion is introduced only when comparing the transformed numerical fields with the analytical results.
In particular, the numerical GG matter function at the center directly determines the
large-$D$ integration function through
\begin{equation}
\label{eq:betaFromNumerics}
\widetilde\beta(\tau)=\frac{\Pi(\tau,0)}{\sqrt{\epsilon}}\qquad \qquad \epsilon=\frac{1}{D-1}\,.
\end{equation}

\subsection{Simplified large-$\boldsymbol{D}$ perturbation theory}

The perturbative expansion of \cite{Ecker:2026akf} simplifies considerably upon using the scaling of the free function $\beta$ discussed in the main text, $\beta(\tau)=\sqrt{\epsilon}\,\tilde\beta(\tau)$. Inserting the series expansions (primes denote derivatives with respect to $\tau$)
\begin{subequations}
    \label{eq:expansions}
\begin{align}
    f &= 1 - \epsilon\,\frac{1-x^2}{2}\,\tilde\beta^2 + \sum_{n=2}^\infty f_n\,\epsilon^n &
    \Omega &= \epsilon\,\tilde\beta^2 + \sum_{n=2}^\infty \Omega_n\,\epsilon^n\\
    \Pi &= \sqrt{\epsilon}^{-1}\,\Big(\epsilon\,\tilde\beta + \sum_{n=2}^\infty \Pi_n\,\epsilon^n\Big) &
    \Psi &= \sqrt{\epsilon}\,\Big(\epsilon\,(\tilde\beta^\prime+\tilde\beta) + \sum_{n=2}^\infty \Psi_n\,\epsilon^n\Big)
\end{align}
\end{subequations}
into the equations of motion in GG gauge,
\begin{align}
x\partial_x\Omega + x^2\Omega^2 &= \frac1\epsilon\,\big((1-2\epsilon)\,\Pi^2-\Omega\big)+ \Omega\,\big(\Pi^2x^2+x^4\,\Psi^2\big)+\Big(\frac1\epsilon-2\Big)\,x^2\,\Psi^2 \label{eq:eom1} \\
   \frac1x\,\partial_x\ln f &=\Omega \label{eq:eom2} \\
   \big(\partial_\tau + x\,\partial_x\big)\,\Pi &= f\,x\,\partial_x \Psi + f\Big(\frac1\epsilon+x^2\,\Omega\Big)\,\Psi-\Pi \label{eq:eom3}\\
   \big(\partial_\tau + x\,\partial_x\big)\,\Psi &= f\,\frac1x\,\partial_x \Pi + f\,\Omega\, \Pi - 2 \Psi \label{eq:eom4}
\end{align}
yields the next-to-leading order (NLO) corrections
\begin{align}
f_2 &= \frac18\,\big(1-x^2\big)\,\tilde\beta\,\big(4(1-x^2)\tilde\beta+(3+x^2)\tilde\beta^3-2(1+x^2)(\tilde\beta''+3\tilde\beta')\big)\\
\Omega_2 &= \tilde\beta\,\big(x^2(\tilde\beta''+3\tilde\beta'-\tilde\beta^3)-2(1-x^2)\tilde\beta\big) \\
\Pi_2 &= \frac{x^2}{2}\,\big(2\tilde\beta+\tilde\beta''+3\tilde\beta'-\tilde\beta^3\big)\\ 
\Psi_2 &=\frac{1}{2}\,\Big(\beta^3+\beta^2\beta'+x^2\big(-4\beta^3-4\beta^2\beta'+6\beta+11\beta'+6\beta''+\beta'''\big)\Big)
\end{align}
and the next-to-next-to-leading order (NNLO) corrections
\begin{align}
f_3 &= c_0+x^2\,c_2+x^4\,c_4+x^6\,c_6\\
\Omega_3 &= 2\beta \Pi_3+\frac{x^4}{4}\,\Big(\beta^6-4\beta^4-6\beta^3\beta'-2\beta^3\beta''+4\beta^2+12\beta\beta'+4\beta\beta''+9(\beta')^2+6\beta'\beta''+(\beta'')^2\Big)\nonumber\\
&\quad+x^2\,\Big(4\beta^4-7\beta^2-10\beta\beta'-4\beta\beta''+(\beta')^2\Big)\\
\Pi_3 &=\frac{x^2}{6}\,\Big(24\beta^3+15\beta^2\beta'+3\beta^2\beta''\Big)+\frac{x^4}{6}\,\Big(3\beta^5-16\beta^3-20\beta^2\beta'-4\beta^2\beta''+12\beta+22\beta'+12\beta''+2\beta'''\Big)\\
\Psi_3 &= \textrm{not\;needed}
\end{align}
with
\begin{align}
c_0 &=
-\frac{17}{48}\beta^6-\frac{79}{36}\beta^4
+\frac{35}{72}\beta^3\beta'
+\frac{13}{72}\beta^3\beta''
+\frac{11}{12}\beta^2
+\frac{7}{9}\beta\beta'
+\frac{1}{6}\beta\beta''
-\frac{1}{9}\beta\beta'''
-\frac{5}{8}(\beta')^2
-\frac{1}{4}\beta'\beta''
-\frac{1}{24}(\beta'')^2
\\
c_2 &=
\frac{3}{16}\beta^6+\frac{3}{4}\beta^4
-\frac{3}{8}\beta^3\beta'
-\frac{1}{8}\beta^3\beta''
\\
c_4 &=
\frac{1}{16}\beta^6+\frac{9}{4}\beta^4
+\frac{7}{8}\beta^3\beta'
+\frac{1}{8}\beta^3\beta''
-\frac{7}{4}\beta^2
-\frac{5}{2}\beta\beta'
-\beta\beta''+\frac{1}{4}(\beta')^2
\\
c_6 &=
\frac{5}{48}\beta^6-\frac{29}{36}\beta^4
-\frac{71}{72}\beta^3\beta'
-\frac{13}{72}\beta^3\beta''
+\frac{5}{6}\beta^2
+\frac{31}{18}\beta\beta'
+\frac{5}{6}\beta\beta''
+\frac{1}{9}\beta\beta'''
+\frac{3}{8}(\beta')^2
+\frac{1}{4}\beta'\beta''
+\frac{1}{24}(\beta'')^2\,.
\end{align}

Note that \eqref{eq:eom4} together with the expansions \eqref{eq:expansions} implies to leading order (LO) $\partial_x\Pi=0$, which explains why $\tilde\beta$ can only depend on $\tau$. The other LO expressions in \eqref{eq:expansions} follow immediately from solving the equations of motion \eqref{eq:eom1}-\eqref{eq:eom4} to LO.

\subsection{Construction of the large-$\boldsymbol{D}$ model}

We summarize here the construction of the large-$D$ model introduced in the main text. Our strategy is to extrapolate the numerical scalar profiles $\Pi_D(\tau,0)$, obtained for dimensions up to $D=200$, to infinite $D$ and identify a simple analytical representation of the limiting function.

We first introduce the rescaled center profile
\begin{equation}
\label{eq:beta_numerical_SM}
\tilde \beta_D(\theta)=\frac{1}{\sqrt{\epsilon}}\,
\Pi_D\left(\frac{\Delta(D)}{2\pi}\theta,0\right)
\end{equation}
where $\theta=2\pi\tau/\Delta(D)$ is the normalized GG time coordinate. In this parametrization, all profiles have period $2\pi$ and satisfy the half-period antisymmetry $\tilde \beta_D(\theta+\pi)=-\tilde \beta_D(\theta)$. Consequently, they admit a Fourier decomposition containing only odd harmonics,
\begin{equation}
\label{eq:beta_fourier_SM}
\tilde \beta_D(\theta)=
\sum_{n=0}^{\infty}
\left[
c_n(D)\cos((2n+1)\theta)
+s_n(D)\sin((2n+1)\theta)
\right]\,.
\end{equation}
For each dimension, the coefficients $c_n(D)$ and $s_n(D)$ are determined by a linear least-squares fit to the numerical profile. Before comparing different dimensions, we fix the arbitrary time origin by aligning the fundamental Fourier harmonics, such that $c_0(D)>0$ and $s_0(D)=0$.

We then assume a regular expansion of the normalized profile in powers of $\epsilon$,
\begin{equation}
\label{eq:beta_expansion_SM}
\tilde \beta_D(\theta)=\tilde \beta_0(\theta)+\epsilon \tilde \beta_1(\theta)+\mathcal O(\epsilon^2)\,.
\end{equation}
To determine the function $\tilde\beta_0$, we independently extrapolate each Fourier coefficient according to
\begin{equation}
c_n(D)=c_{n,0}+c_{n,1}\epsilon
+c_{n,2}\epsilon^2+\cdots\qquad\qquad
s_n(D)=s_{n,0}+s_{n,1}\epsilon
+s_{n,2}\epsilon^2+\cdots\,.
\end{equation}
The leading-order profiles are therefore reconstructed as
\begin{equation}
\label{eq:beta_orders_SM}
\tilde \beta_0(\theta)=
\sum_{n=0}^{\infty}
\left[
c_{n,0}\cos((2n+1)\theta)
+s_{n,0}\sin((2n+1)\theta)
\right]
\end{equation}
This procedure determines $\tilde\beta_0(\theta)$ from the numerical solutions, without imposing a particular analytical ansatz for its shape.

To identify a simpler representation of $\tilde\beta_0$, we express its Fourier coefficients in terms of amplitudes and phases,
\begin{equation}
A_n=\sqrt{c_{n,0}^2+s_{n,0}^2} \qquad\qquad \varphi_n =-2\arctan\frac{s_{n,0}}{A_n+c_{n,0}}\,.
\end{equation}
We find that the amplitudes decrease approximately geometrically, $A_n\simeq A_*q^n$, while the phases exhibit an approximately linear progression, $\varphi_n\simeq n(\pi+\delta)$, modulo $2\pi$. These observations motivate the ansatz
\begin{equation}
\label{eq:beta_toy_SM}
\tilde \beta(\theta)=A_*\sum_{n=0}^{\infty}(-q)^n\cos[(2n+1)\theta+n\delta]=A_*\frac{\cos\theta+q\cos(\theta+\delta)}{1+2q\cos(2\theta+\delta)+q^2}=A_\ast\,\operatorname{Re}\left[\frac{e^{i\theta}}{1+q e^{i(2\theta+\delta)}}\right]
\end{equation}
which provides a smooth, closed-form approximation to the limiting numerical profile $\tilde \beta_0(\theta)$ for $0<q<1$.

Importantly, the normalization of this ansatz is not an independent fitting parameter. As derived in the main text, the asymptotic expansions of $\mu$ and $\Delta$ imply
\begin{equation}
\label{eq:beta_norm_SM}
\mathcal N_0\equiv
\frac{\Delta_1-4\pi\mu_1}{4\pi\mu_0}
=
\frac{1}{2\pi}\int_0^{2\pi}
\tilde \beta_0(\theta)^2\,\mathrm d\theta.
\end{equation}
For the toy model \eqref{eq:beta_toy_SM}, orthogonality of the Fourier modes gives
\begin{equation}
\frac{1}{2\pi}\int_0^{2\pi}
\tilde \beta(\theta)^2\,\mathrm d\theta
=
\frac{A_*^2}{2(1-q^2)}.
\end{equation}
Imposing the analytical normalization therefore fixes
\begin{equation}
\label{eq:beta_amplitude_SM}
A_*=\sqrt{2\mathcal N_0(1-q^2)}.
\end{equation}
The model consequently contains only two free shape parameters, $q$ and $\delta$. We determine these by minimizing the squared differences between the first four Fourier coefficients ($c_{n,0}$, $s_{n,0}$) and those predicted by the ansatz ($\bar c_{n}$, $\bar s_{n}$),
\begin{equation}
\label{eq:beta_fit_SM}
\chi^2(q,\delta)=
\sum_{n=0}^{3}
\left[
(c_{n,0}-\bar c_n)^2+
(s_{n,0}-\bar s_n)^2
\right]\qquad \bar c_n=A_*(-q)^n\cos(n\delta)\qquad
\bar s_n=-A_*(-q)^n\sin(n\delta)\,,
\end{equation}
where $A_*$ is determined by Eq.~\eqref{eq:beta_amplitude_SM} for each trial value of $q$. 
With this procedure we arrive at the results for the shape parameters $q\approx0.0551$ and $\delta\approx0.0985$. Substituting this approximation into the large-$D$ expressions for the metric and matter fields yields the center profiles compared with the numerical solutions in Fig.~\ref{fig:4} of the main text.

\begin{figure}[htb]
\centering
\includegraphics[width=0.92\linewidth]{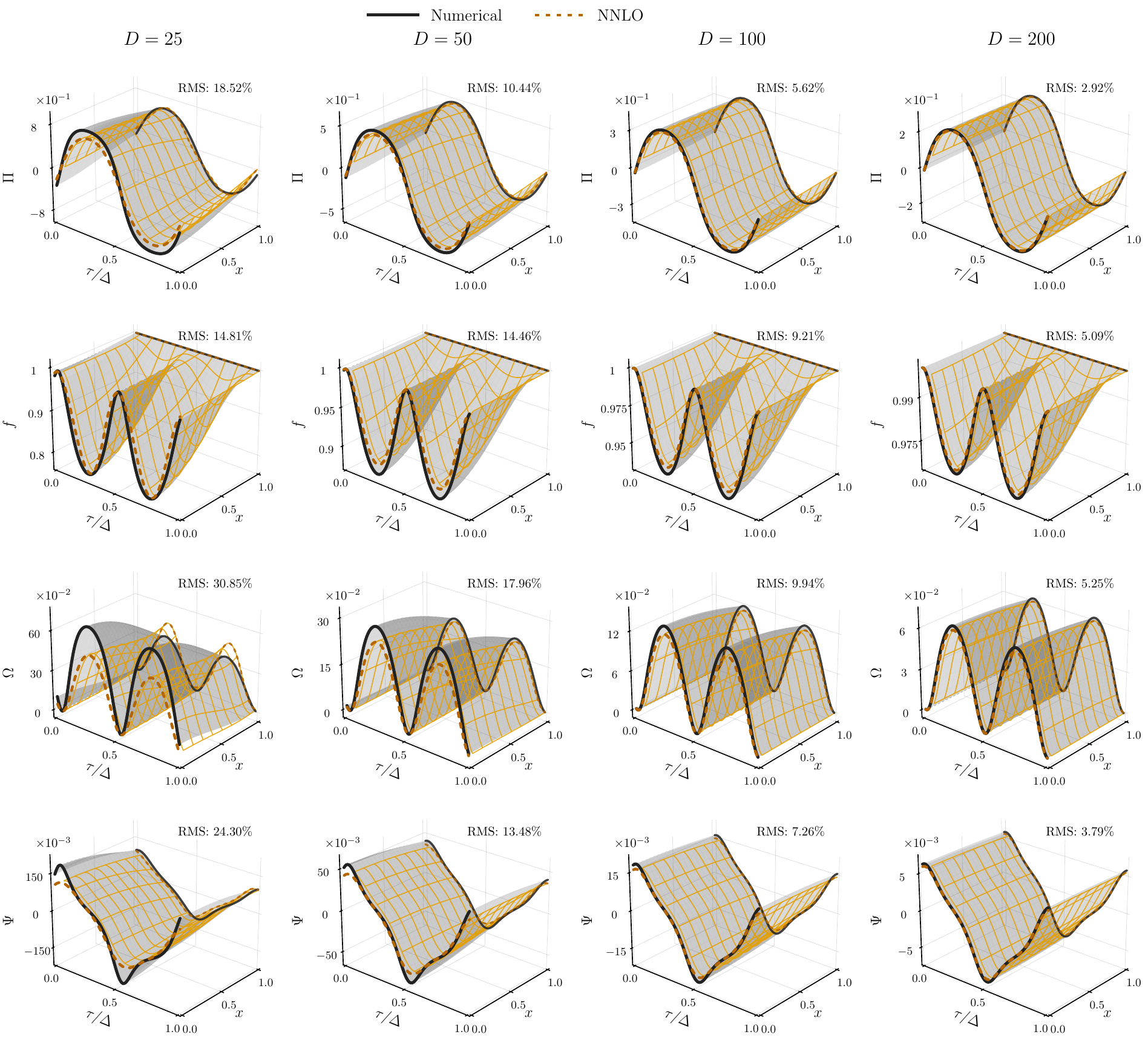}
\caption{Comparison of the numerical solutions (gray surfaces) with the NNLO predictions (orange dotted meshes) of the large-$D$ model for $D=25,50,100$, and $200$.
The relative RMS deviations over the $(\tau,x)$ domain are indicated in each panel. }
\label{fig:6}
\end{figure}

To assess the validity of our large-$D$ model, Fig.~\ref{fig:6} compares the numerical solutions for $D=25,50,100$, and $200$ with the NNLO perturbative predictions over the full $(\tau,x)$ domain.
The relative root-mean-square (RMS) deviations, indicated in each panel, decrease systematically with increasing $D$ for all four fields, $\Pi$, $f$, $\Omega$, and $\Psi$, providing numerical evidence that the finite-$D$ solutions indeed converge toward the proposed large-$D$ model.

\begin{table*}[p]
\centering
\caption{Echoing period $\Delta$, computed in Ref.~\cite{Ecker:2026joz} for $3.046 \leq D \leq 5.5$ and in this work for all other dimensions.}
\label{tab:1}
\begingroup
\setlength{\tabcolsep}{1.5pt}
\renewcommand{\arraystretch}{0.98}
\fontsize{7.8}{9.1}\selectfont
\begin{tabular}{@{}r@{\hspace{0.40em}}r@{\hspace{0.90em}}r@{\hspace{0.40em}}r@{\hspace{0.90em}}r@{\hspace{0.40em}}r@{\hspace{0.90em}}r@{\hspace{0.40em}}r@{\hspace{0.90em}}r@{\hspace{0.40em}}r@{\hspace{0.90em}}r@{\hspace{0.40em}}r@{\hspace{0.90em}}r@{\hspace{0.40em}}r@{}}
\toprule
$D$ & $\Delta$ & $D$ & $\Delta$ & $D$ & $\Delta$ & $D$ & $\Delta$ & $D$ & $\Delta$ & $D$ & $\Delta$ & $D$ & $\Delta$ \\
\midrule
3.046 & 2.454367 & 3.112 & 2.825094 & 4.35 & 3.376097 & 4.95 & 3.233510 & 12 & 2.389189 & 58 & 1.815397 & 130 & 1.722127 \\
3.0465 & 2.458274 & 3.113 & 2.829191 & 4.39 & 3.366930 & 4.955 & 3.232331 & 12.25 & 2.375503 & 59 & 1.812598 & 131 & 1.721541 \\
3.047 & 2.462156 & 3.114 & 2.833257 & 4.4 & 3.364618 & 4.96 & 3.231153 & 12.5 & 2.362324 & 60 & 1.809890 & 132 & 1.720964 \\
3.0475 & 2.466012 & 3.115 & 2.837291 & 4.41 & 3.362298 & 4.965 & 3.229976 & 12.75 & 2.349623 & 61 & 1.807267 & 133 & 1.720395 \\
3.048 & 2.469844 & 3.116 & 2.841294 & 4.42 & 3.359971 & 4.97 & 3.228800 & 13 & 2.337375 & 62 & 1.804726 & 134 & 1.719835 \\
3.0485 & 2.473651 & 3.117 & 2.845267 & 4.43 & 3.357638 & 4.975 & 3.227625 & 13.25 & 2.325554 & 63 & 1.802263 & 135 & 1.719283 \\
3.049 & 2.477433 & 3.118 & 2.849209 & 4.44 & 3.355298 & 4.98 & 3.226450 & 13.5 & 2.314139 & 64 & 1.799875 & 136 & 1.718739 \\
3.0495 & 2.481192 & 3.119 & 2.853122 & 4.45 & 3.352953 & 4.985 & 3.225277 & 13.75 & 2.303109 & 65 & 1.797557 & 137 & 1.718202 \\
3.05 & 2.484927 & 3.12 & 2.857005 & 4.46 & 3.350601 & 4.99 & 3.224104 & 14 & 2.292445 & 66 & 1.795308 & 138 & 1.717674 \\
3.0505 & 2.488639 & 3.121 & 2.860860 & 4.47 & 3.348245 & 4.995 & 3.222932 & 14.25 & 2.282128 & 67 & 1.793124 & 139 & 1.717152 \\
3.051 & 2.492328 & 3.122 & 2.864685 & 4.48 & 3.345883 & 5 & 3.221761 & 14.5 & 2.272142 & 68 & 1.791002 & 140 & 1.716638 \\
3.0515 & 2.495994 & 3.123 & 2.868482 & 4.49 & 3.343517 & 5.005 & 3.220591 & 14.75 & 2.262470 & 69 & 1.788940 & 141 & 1.716132 \\
3.052 & 2.499638 & 3.124 & 2.872251 & 4.5 & 3.341147 & 5.025 & 3.215920 & 15 & 2.253098 & 70 & 1.786935 & 142 & 1.715632 \\
3.0525 & 2.503259 & 3.125 & 2.875992 & 4.51 & 3.338772 & 5.045 & 3.211263 & 15.5 & 2.235199 & 71 & 1.784985 & 143 & 1.715139 \\
3.053 & 2.506858 & 3.126 & 2.879706 & 4.52 & 3.336394 & 5.065 & 3.206622 & 16 & 2.218343 & 72 & 1.783087 & 144 & 1.714653 \\
3.054 & 2.513993 & 3.127 & 2.883393 & 4.53 & 3.334012 & 5.085 & 3.201996 & 16.5 & 2.202440 & 73 & 1.781240 & 145 & 1.714173 \\
3.055 & 2.521043 & 3.128 & 2.887052 & 4.54 & 3.331628 & 5.105 & 3.197387 & 17 & 2.187410 & 74 & 1.779441 & 146 & 1.713700 \\
3.056 & 2.528010 & 3.129 & 2.890685 & 4.55 & 3.329240 & 5.125 & 3.192795 & 17.5 & 2.173183 & 75 & 1.777689 & 147 & 1.713233 \\
3.057 & 2.534896 & 3.13 & 2.894292 & 4.56 & 3.326849 & 5.135 & 3.190505 & 18 & 2.159695 & 76 & 1.775982 & 148 & 1.712773 \\
3.058 & 2.541704 & 3.131 & 2.897873 & 4.57 & 3.324456 & 5.145 & 3.188219 & 18.25 & 2.153211 & 77 & 1.774318 & 149 & 1.712318 \\
3.059 & 2.548434 & 3.132 & 2.901428 & 4.58 & 3.322061 & 5.149 & 3.187306 & 18.5 & 2.146890 & 78 & 1.772695 & 150 & 1.711870 \\
3.06 & 2.555090 & 3.133 & 2.904958 & 4.59 & 3.319664 & 5.151 & 3.186850 & 18.75 & 2.140727 & 79 & 1.771113 & 151 & 1.711427 \\
3.061 & 2.561671 & 3.134 & 2.908462 & 4.6 & 3.317265 & 5.153 & 3.186394 & 19 & 2.134715 & 80 & 1.769568 & 152 & 1.710990 \\
3.062 & 2.568180 & 3.135 & 2.911942 & 4.61 & 3.314865 & 5.155 & 3.185938 & 19.5 & 2.123126 & 81 & 1.768061 & 153 & 1.710559 \\
3.063 & 2.574618 & 3.14 & 2.928976 & 4.62 & 3.312463 & 5.157 & 3.185482 & 20 & 2.112080 & 82 & 1.766590 & 154 & 1.710133 \\
3.064 & 2.580987 & 3.145 & 2.945426 & 4.63 & 3.310061 & 5.159 & 3.185027 & 20.5 & 2.101540 & 83 & 1.765153 & 155 & 1.709713 \\
3.065 & 2.587288 & 3.15 & 2.961320 & 4.64 & 3.307657 & 5.169 & 3.182752 & 20.5625 & 2.100256 & 84 & 1.763750 & 156 & 1.709298 \\
3.066 & 2.593522 & 3.155 & 2.976684 & 4.65 & 3.305253 & 5.179 & 3.180481 & 20.75 & 2.096449 & 85 & 1.762379 & 157 & 1.708888 \\
3.067 & 2.599691 & 3.16 & 2.991545 & 4.66 & 3.302849 & 5.189 & 3.178215 & 21 & 2.091485 & 86 & 1.761038 & 158 & 1.708483 \\
3.068 & 2.605796 & 3.165 & 3.005924 & 4.67 & 3.300444 & 5.199 & 3.175954 & 21.5 & 2.081984 & 87 & 1.759728 & 159 & 1.708083 \\
3.069 & 2.611839 & 3.17 & 3.019843 & 4.68 & 3.298039 & 5.209 & 3.173697 & 22 & 2.073239 & 88 & 1.758447 & 160 & 1.707688 \\
3.07 & 2.617820 & 3.175 & 3.033323 & 4.69 & 3.295634 & 5.219 & 3.171444 & 22.5 & 2.065799 & 89 & 1.757194 & 161 & 1.707298 \\
3.071 & 2.623741 & 3.18 & 3.046381 & 4.7 & 3.293229 & 5.229 & 3.169197 & 23 & 2.055322 & 90 & 1.755968 & 162 & 1.706913 \\
3.072 & 2.629602 & 3.2 & 3.094740 & 4.71 & 3.290825 & 5.239 & 3.166954 & 23.5 & 2.047598 & 91 & 1.754768 & 163 & 1.706532 \\
3.073 & 2.635405 & 3.25 & 3.193154 & 4.72 & 3.288421 & 5.249 & 3.164716 & 24 & 2.039859 & 92 & 1.753594 & 164 & 1.706156 \\
3.074 & 2.641151 & 3.3 & 3.267213 & 4.73 & 3.286018 & 5.259 & 3.162483 & 24.5 & 2.032430 & 93 & 1.752445 & 165 & 1.705785 \\
3.075 & 2.646841 & 3.35 & 3.323462 & 4.74 & 3.283616 & 5.269 & 3.160254 & 25 & 2.025294 & 94 & 1.751319 & 166 & 1.705418 \\
3.076 & 2.652476 & 3.4 & 3.366252 & 4.75 & 3.281215 & 5.279 & 3.158031 & 25.5 & 2.018435 & 95 & 1.750217 & 167 & 1.705055 \\
3.077 & 2.658056 & 3.45 & 3.398625 & 4.76 & 3.278815 & 5.319 & 3.149185 & 26 & 2.011840 & 96 & 1.749137 & 168 & 1.704696 \\
3.078 & 2.663583 & 3.5 & 3.422791 & 4.77 & 3.276417 & 5.349 & 3.142604 & 26.5 & 2.005494 & 97 & 1.748078 & 169 & 1.704342 \\
3.079 & 2.669058 & 3.55 & 3.440402 & 4.78 & 3.274020 & 5.359 & 3.140421 & 27 & 1.999387 & 98 & 1.747041 & 170 & 1.703992 \\
3.08 & 2.674482 & 3.6 & 3.452721 & 4.79 & 3.271625 & 5.369 & 3.138242 & 27.5 & 1.992233 & 99 & 1.746024 & 171 & 1.703646 \\
3.081 & 2.679855 & 3.65 & 3.460729 & 4.8 & 3.269188 & 5.379 & 3.136069 & 28 & 1.986410 & 100 & 1.745027 & 172 & 1.703303 \\
3.082 & 2.685178 & 3.7 & 3.465201 & 4.805 & 3.267991 & 5.389 & 3.133901 & 29 & 1.975334 & 101 & 1.744050 & 173 & 1.702965 \\
3.083 & 2.690452 & 3.71 & 3.465727 & 4.81 & 3.266794 & 5.409 & 3.129581 & 30 & 1.964958 & 102 & 1.743091 & 174 & 1.702631 \\
3.084 & 2.695678 & 3.72 & 3.466142 & 4.815 & 3.265597 & 5.419 & 3.127429 & 31 & 1.955218 & 103 & 1.742150 & 175 & 1.702300 \\
3.085 & 2.700856 & 3.73 & 3.466449 & 4.82 & 3.264401 & 5.459 & 3.118874 & 32 & 1.946056 & 104 & 1.741228 & 176 & 1.701973 \\
3.086 & 2.705988 & 3.74 & 3.466653 & 4.825 & 3.263205 & 5.489 & 3.112515 & 33 & 1.937423 & 105 & 1.740322 & 177 & 1.701649 \\
3.087 & 2.711074 & 3.75 & 3.466756 & 4.83 & 3.262010 & 5.499 & 3.110407 & 34 & 1.929273 & 106 & 1.739433 & 178 & 1.701330 \\
3.088 & 2.716115 & 3.755 & 3.466771 & 4.835 & 3.260815 & 5.5 & 3.110196 & 35 & 1.921567 & 107 & 1.738560 & 179 & 1.701013 \\
3.089 & 2.721111 & 3.7554 & 3.466772 & 4.84 & 3.259621 & 5.75 & 3.058068 & 36 & 1.914268 & 108 & 1.737703 & 180 & 1.700701 \\
3.09 & 2.726063 & 3.7555 & 3.466772 & 4.845 & 3.258427 & 6 & 3.009483 & 37 & 1.907347 & 109 & 1.736862 & 181 & 1.700391 \\
3.091 & 2.730973 & 3.7556 & 3.466772 & 4.85 & 3.257234 & 6.25 & 2.963826 & 38 & 1.900773 & 110 & 1.736035 & 182 & 1.700085 \\
3.092 & 2.735839 & 3.7557 & 3.466772 & 4.855 & 3.256042 & 6.5 & 2.920939 & 39 & 1.894521 & 111 & 1.735223 & 183 & 1.699782 \\
3.093 & 2.740664 & 3.7558 & 3.466772 & 4.86 & 3.254850 & 6.75 & 2.880644 & 40 & 1.888568 & 112 & 1.734425 & 184 & 1.699483 \\
3.094 & 2.745447 & 3.7559 & 3.466772 & 4.865 & 3.253659 & 7 & 2.842753 & 41 & 1.882893 & 113 & 1.733642 & 185 & 1.699187 \\
3.095 & 2.750189 & 3.756 & 3.466772 & 4.87 & 3.252468 & 7.5 & 2.773476 & 42 & 1.877477 & 114 & 1.732871 & 186 & 1.698893 \\
3.096 & 2.754892 & 3.76 & 3.466763 & 4.875 & 3.251278 & 8 & 2.711790 & 43 & 1.872302 & 115 & 1.732114 & 187 & 1.698603 \\
3.097 & 2.759555 & 3.77 & 3.466677 & 4.88 & 3.250089 & 8.5 & 2.656569 & 44 & 1.867353 & 116 & 1.731369 & 188 & 1.698316 \\
3.098 & 2.764178 & 3.78 & 3.466502 & 4.885 & 3.248900 & 8.75 & 2.631083 & 45 & 1.862614 & 117 & 1.730637 & 189 & 1.698032 \\
3.099 & 2.768763 & 3.79 & 3.466241 & 4.89 & 3.247712 & 9 & 2.606874 & 46 & 1.858074 & 118 & 1.729917 & 190 & 1.697751 \\
3.1 & 2.773311 & 3.8 & 3.465896 & 4.895 & 3.246524 & 9.25 & 2.583850 & 47 & 1.853719 & 119 & 1.729209 & 191 & 1.697473 \\
3.101 & 2.777821 & 3.85 & 3.463028 & 4.9 & 3.245338 & 9.5 & 2.561926 & 48 & 1.849538 & 120 & 1.728513 & 192 & 1.697198 \\
3.103 & 2.786730 & 3.9 & 3.458488 & 4.905 & 3.244151 & 9.75 & 2.541026 & 49 & 1.845522 & 121 & 1.727828 & 193 & 1.696925 \\
3.104 & 2.791130 & 3.95 & 3.452553 & 4.91 & 3.242966 & 10 & 2.521079 & 50 & 1.841660 & 122 & 1.727154 & 194 & 1.696656 \\
3.105 & 2.795495 & 4 & 3.445453 & 4.915 & 3.241781 & 10.25 & 2.502022 & 51 & 1.837943 & 123 & 1.726490 & 195 & 1.696389 \\
3.106 & 2.799825 & 4.05 & 3.437381 & 4.92 & 3.240597 & 10.5 & 2.483796 & 52 & 1.834365 & 124 & 1.725837 & 196 & 1.696124 \\
3.107 & 2.804121 & 4.1 & 3.428501 & 4.925 & 3.239414 & 10.75 & 2.466348 & 53 & 1.830916 & 125 & 1.725195 & 197 & 1.695863 \\
3.108 & 2.808382 & 4.15 & 3.418948 & 4.93 & 3.238232 & 11 & 2.449629 & 54 & 1.827591 & 126 & 1.724562 & 198 & 1.695604 \\
3.109 & 2.812609 & 4.2 & 3.408840 & 4.935 & 3.237050 & 11.25 & 2.433593 & 55 & 1.824382 & 127 & 1.723939 & 199 & 1.695347 \\
3.11 & 2.816804 & 4.25 & 3.398275 & 4.94 & 3.235869 & 11.5 & 2.418200 & 56 & 1.821284 & 128 & 1.723326 & 200 & 1.695093 \\
3.111 & 2.820965 & 4.3 & 3.387336 & 4.945 & 3.234689 & 11.75 & 2.403410 & 57 & 1.818291 & 129 & 1.722722 & $\infty$ & 1.643911 \\
\bottomrule
\end{tabular}
\endgroup
\end{table*}

\end{document}